\documentclass{aa} 

\bibpunct{(}{)}{;}{a}{}{,} 

\usepackage[varg]{txfonts}
\usepackage{amssymb}
\usepackage{amsmath}
\usepackage{natbib}
\usepackage{subcaption}
\usepackage{graphicx}
\usepackage{epstopdf}
\usepackage{marvosym}
\usepackage{epsfig}
\usepackage{multirow}
\usepackage{tikz}
\usepackage{nicefrac}
\usepackage{arydshln}
\usepackage{url}
\usepackage{blkarray}
\usepackage{hyphenat} 
\usepackage{lipsum}
\usepackage{caption}
\usepackage{booktabs}
\usetikzlibrary{arrows,calc,intersections,positioning,arrows,decorations.pathmorphing,decorations.markings,shapes}

\newcommand{\RN}[1]{
  \textup{\uppercase\expandafter{\romannumeral#1}}
}
\begin{document}

\title{Analytical framework for mutual approximations}
\subtitle{Derivation and application to Jovian satellites}

\author{M. Fayolle \inst{\ref{TUDELFT}}
	\and D. Dirkx \inst{\ref{TUDELFT}} \and P.N.A.M. Visser \inst{\ref{TUDELFT}} \and V. Lainey \inst{\ref{IMCCE}} }

\institute{Delft University of Technology, Kluyverweg 1, 2629HS Delft, The Netherlands \\ \email{m.s.fayolle-chambe@tudelft.nl} \label{TUDELFT}
\and IMCCE, Observatoire de Paris, PSL Research University, CNRS-UMR8028 du CNRS, UPMC, Lille-1, 77 Av. Denfert-Rochereau, 75014, Paris, France \label{IMCCE}}

\date{Received 7 May 2021 / Accepted 1 June 2021}

\abstract
{The apparent close encounters of two satellites in the plane of the sky, called mutual approximations, have been suggested as a different type of astrometric observation to refine the moons' ephemerides. The main observables are then the central instants of the close encounters, which have the advantage of being free of any scaling and orientation errors. However, no analytical formulation is available yet for the observation partials of these central instants, leaving numerical approaches or alternative observables (i.e. derivatives of the apparent distance instead of central instants) as options.}
{Filling that gap, this paper develops an analytical method to include central instants as direct observables in the ephemerides estimation and assesses the quality of the resulting solution.}
{To this end, the apparent relative position between the two satellites is approximated by a second-order polynomial near the close encounter. This eventually leads to an expression for mutual approximations' central instants as a function of the apparent relative position, velocity, and acceleration between the two satellites.}
{The resulting analytical expressions for the central instant partials were validated numerically. In addition, we ran a covariance analysis to compare the estimated solutions obtained with the two types of observables (central instants versus alternative observables), using the Galilean moons of Jupiter as a test case. Our analysis shows that alternative observables are almost equivalent to central instants in most cases. Accurate individual weighting of each alternative observable, accounting for the mutual approximation's characteristics (which are automatically included in the central instants' definition), is however crucial to obtain consistent solutions between the two observable types. Using central instants still yields a small improvement of 10-20\% of the formal errors in the radial and normal directions (RSW frame), compared to the alternative observables' solution. This improvement increases when mutual approximations with low impact parameters and large impact velocities are included in the estimation.}
{Choosing between the two observables thus requires careful assessment, taking into account the characteristics of the available observations. Using central instants over alternative observables ensures that the state estimation fully benefits from the information encoded in mutual approximations, which might be necessary depending on the application of the ephemeris solution.}


\maketitle

\section{Introduction} \label{sec:introduction}

Natural satellites are among the most fascinating objects in our Solar System. In particular, leading candidates for extraterrestrial habitats are found among Jovian and Saturnian satellites. Knowing more about the past history of these moons is key to understanding whether they offer life-favourable conditions now and, therefore, to analysing the conditions for habitability in our Solar System and beyond \citep{marion2003search,parkinson2008habitability,lunine2017ocean}. However, the moons' origin and evolution still remain poorly understood, while they are crucial to investigate the existence and stability of these putative habitats \citep[e.g.][]{crida2012formation,cuk2016dynamical,fuller2016resonance}. 

Measuring and fitting the current motion of natural satellites provides valuable insights into their dynamical history. In particular, it helps to understand tidal dissipation mechanisms, which play a crucial role in planetary systems' orbital evolution \citep{lainey2009strong,lainey2012strong,lainey2020resonance,fuller2016resonance}. More generally, determining natural satellites' dynamics indirectly gives hints about planetary formation processes \citep[e.g.][]{heller2015formation,samuel2019rheology}.  

As our interest in natural satellites grows, more dedicated missions are being proposed to explore them (JUICE, Europa-Clipper, IVO, MMX, etc.). Precise knowledge of the moons' current states then also becomes crucial to optimise the orbital design of such missions, for instance to propose efficient orbital insertions and flybys \citep{murrow1988galilean,raofi2000preliminary,lynam2012preliminary}. Due to inaccuracies in the predicted state of the targeted body, corrective manoeuvres are indeed required before and after flybys (or, similarly, orbital insertions) and can be significantly reduced by improved ephemerides.

Determining the orbits of natural satellites is typically achieved with observations of their absolute positions in the sky or of their relative motion with respect to one another.  Spacecraft-based observations (either radiometric tracking or optical data) can also be used, but they are much sparser because they are only collected during planetary missions. 

Extremely precise measurements are necessary to be sensitive to very weak dynamical effects, such as tidal forces, which drive the orbital evolution of planetary systems. 
Unfortunately, the precision of Earth-based classical astrometric observations is limited, typically ranging from 50 to 150 milliarcseconds (mas) \citep[e.g.][]{stone2001positions,kiseleva2008astrometric,robert2017ccd}.

A lot of effort has thus been dedicated to develop more precise types of observations. For example, relative measurements of the positions of two satellites in the sky plane have been shown to be more accurate, with a precision down to 30 mas \citep{peng2012precise}. Relative astrometric observations can indeed benefit from the so-called precision premium: the precision is significantly improved when apparent distances get smaller than 85 mas. In such a situation, instrumental and astronomical error sources tend to have a similar effect on the measurement of each of the two satellites' position, and thus they eventually cancel out \citep{pascu1994galactic,peng2008ccd}. 

Alternatively, the relative position of two satellites can also be precisely measured by observing mutual events - occultations or eclipses \citep[e.g.][]{emelyanov2009mutual,emelyanov2011astrometric,dias2013analysis,arlot2014phemu09}. During mutual events, one satellite masks the other, resulting in a drop of the flux received by the observer. Those mutual phenomena can provide measurements of satellites' relative positions with a precision of about 10 mas \citep{emelyanov2009mutual,dias2013analysis}. However, they can only be witnessed during the equinox of the central planet, which occurs every 6 years for Jupiter and 15 years for Saturn. This significantly limits the number of available observations.

\begin{figure}
	\centering
	\resizebox{\hsize}{!}{\includegraphics{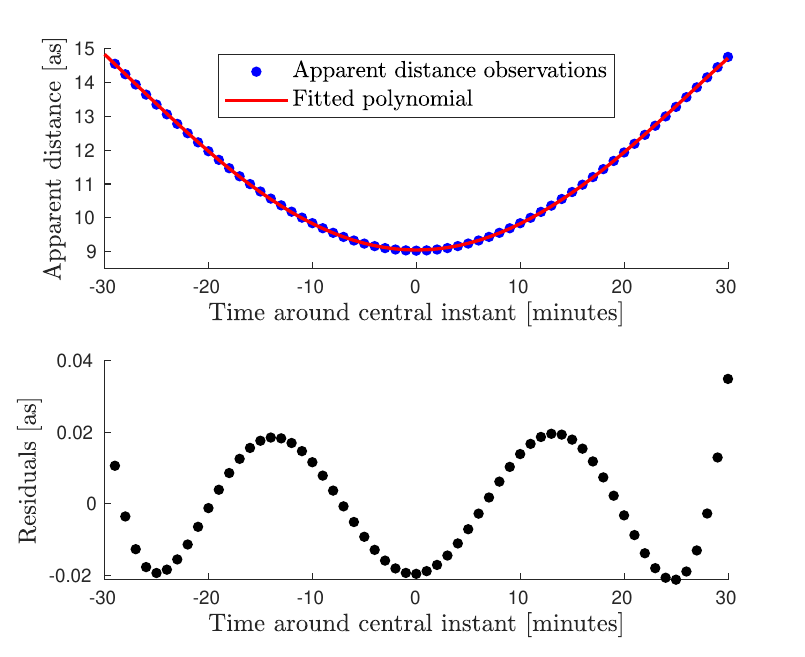}}
	\caption{Observation of a mutual approximation (i.e. close encounter between two natural satellites). The apparent distance between the two satellites (blue dots in the top panel) is frequently measured and a polynomial is used to fit these observations and estimate the central instant of the close encounter (typically fourth-order polynomial, displayed in red). The residuals between the apparent distances measurements and the fitted polynomial are shown in black (bottom panel).}
	\label{fig:mutualApproxDefinition}
\end{figure}

To overcome the limitations of the above-mentioned observations, a very promising alternative technique called mutual approximation was recently proposed by \cite{morgado2016astrometry}, though initially suggested in \cite{arlot1982results}. This method determines the so-called central instant at which a close encounter occurs in the sky plane (i.e. the apparent distance between two satellites reaches a minimum, see Figure \ref{fig:mutualApproxDefinition}). The precision of mutual approximations was found to be comparable to that of mutual events \citep{morgado2016astrometry,morgado2019approx}. 

Central instants are free of any orientation and scaling errors in the instrumental frame: they do not depend on the absolute value of the apparent distance itself, nor on the relative orientation of the two satellites \citep{emelyanov2017precision}. This eliminates two major error sources present in classical astrometric observations. Properly recording the observational time at the ground station becomes crucial, but this can be easily achieved with GPS receivers or dedicated software. Most importantly, mutual approximations occur very regularly, and thus offer a very promising alternative to eclipses and occultations \citep{morgado2016astrometry,morgado2019approx}.

To estimate ephemerides using mutual approximations, the observation partials for central instants are required. They link a small variation of the parameters to be estimated (natural satellites' states in our case) to a change in the observable. However, the central instants' complex definition and their relation to the satellites' states makes deriving these equations difficult. Other astronomic observables only depend on the apparent (relative) position of the observed body which is an indirect function of its inertial position, after projection on the plane of the sky. Mutual approximations, on the other hand, are also determined by the apparent relative velocity and acceleration of the two satellites. As a consequence, such observations are affected by the satellites' inertial relative dynamics, and not only by their position.

\cite{emelyanov2017precision} and \cite{morgado2019approx} therefore assumed that variational equations could not be solved analytically when using central instants, as analytical partials were not yet available (or easily derivable) for such observables. 
Those central instants partials could be computed numerically, but this process is highly computationally demanding \citep{emelyanov2017precision} and can also be error prone. 
Consequently, it was suggested to use a modified observable and fit the derivative of the apparent distance instead of the central instant itself \citep[][Lainey, in preparation]{emelyanov2017precision,morgado2019approx}. This modified observable can be expressed as a simple function of the relative position and velocity of the two satellites (see Section \ref{sec:alternativeObservables}). Moreover, the apparent distance derivative is by definition equal to zero at closest encounter, which significantly simplifies the equations.

This indirect method is currently the recommended approach to obtain the mutual approximations' observation partials \citep{emelyanov2017precision,morgado2019approx}. Fundamentally, defining the central instant $t_c$ directly or stating that the derivative of the apparent distance should be equal to zero at $t_c$ both express the fact that the point of closest approach is reached at this instant. However, the information both observable types convey to the state estimation is not necessarily identical and it has not yet been proven that fitting the derivative of the apparent distance is equivalent to fitting the central instants. Actually, using numerical partials for central instants led to convergence issues in \cite{emelyanov2017precision}, while none were encountered with alternative observables. This would indicate that the two observables are not completely interchangeable.

To extend the current framework available for the mutual approximation technique, this paper develops an analytical formulation for the observation partials of the central instants. To achieve this, the relative motion of the two satellites in the plane of the sky is approximated by a polynomial function around the close encounter. The polynomial coefficients are defined from the relative position, velocity, and acceleration of the two satellites, as seen from the observer. It thus becomes possible to derive analytical expressions for the change in central instant induced by a variation in either the two satellites' or the observer's states. We successfully performed the state estimation with mutual approximations' suggested alternative observables (derivatives of the apparent distance) and with central instants separately, using our analytical observation partials for the latter. This comparison aims at quantifying the influence of the observable choice on the estimated solution. We show that it is essential to adopt an appropriate weighting strategy when using alternative observables to achieve consistent results between the two observable types, but that central instants can nonetheless yield a 10-20\% reduction in formal errors.

We develop the analytical framework for mutual approximations' central instants in Section \ref{sec:formulation}, while the details of the observables simulation and state estimation are provided in Section \ref{sec:stateEstimationProcedure}. The results of our comparative analysis are discussed in Section \ref{sec:results}, first using a simple test case limited to mutual approximations between Io and Europa, before extending it to the four Galilean moons. The main concluding points are summarised in Section \ref{sec:conclusions}. 
All the numerical simulations presented in this paper were conducted using the Tudat toolkit developed by the Astrodynamics \& Space Missions department of Delft University of Technology \citep[see Appendix C in][]{dirkx2019propagation}. 

\section{Using mutual approximations in the estimation} \label{sec:formulation}

In this section, we first provide a formal definition to describe the observation of a mutual approximation between two satellites in Section \ref{sec:mutualApproximationDefinition}. We develop an analytical formulation for the central instants and their observation partials in Section \ref{sec:expressionCentralInstants} and Section \ref{sec:centralInstantPartials}, respectively. The light-time effect contribution to those partials is discussed in Section \ref{sec:lightTimeCorrections}. Finally, the alternative mutual approximations' observable (i.e. derivative of the apparent distance, as introduced in Section \ref{sec:introduction}) is presented in more details in Section \ref{sec:alternativeObservables}.

\subsection{Mutual approximation definition} \label{sec:mutualApproximationDefinition}

A mutual approximation involves an observer (denoted by the subscript $O$ in the following), which is most commonly a ground station, and two natural satellites, between which a close encounter is observed (subscripts $S1$ and $S2$, respectively). Because the light has a finite speed, the time at which the mutual approximation is observed (observation time $t_{_O}$) differs from the time at which the light eventually received by the observer got reflected by each of the satellites ($t_{_{S1}}$ and $t_{_{S2}}$ for satellites 1 and 2, respectively). 

\begin{figure}
	\centering
	\resizebox{\hsize}{!}{\includegraphics{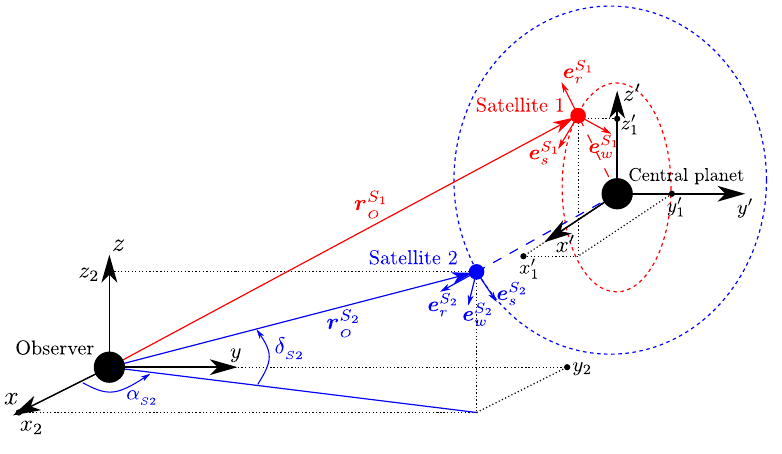}}
	\caption{Schematic representation of the different coordinate systems and positions. The first satellite and all associated notations are depicted in red, while blue is used for the second satellite. $\boldsymbol{r}_{_O}^{S_i}$ denotes the relative position vector between satellite $i$ and the observer, and $\left[x_{i},y_{i},z_{i}\right]$ correspond to the observer-centred cartesian coordinates of satellite $i$. $\alpha_{_{Si}}$ and $\delta_{_{Si}}$ refer to the right ascension and declination of satellite $i$, as seen by the observer. $\left[x^{\prime}_{i},y^{\prime}_{i},z^{\prime}_{i}\right]$ are the satellites' central body-centred cartesian coordinates. $\left[\boldsymbol{e}_{_r}^{S_i},\boldsymbol{e}_{_s}^{S_i},\boldsymbol{e}_{_w}^{S_i}\right]$ defines the RSW reference frame associated with satellite $i$. The vectors $\boldsymbol{e}_{_r}^{S_i}$, $\boldsymbol{e}_{_s}^{S_i}$, and $\boldsymbol{e}_{_w}^{S_i}$ correspond to the radial, normal, and axial directions, respectively.}
	\label{fig:referenceFrames}
\end{figure}

The relative range vectors between the satellites and the observer can thus be defined as follows (see Figure \ref{fig:referenceFrames}):
\begin{align}
\boldsymbol{r}_{_O}^{S_1} &= \boldsymbol{r}_{_{S1}}(t_{_{S1}}) - \boldsymbol{r}_{_O}(t_{_O}),\\
\boldsymbol{r}_{_O}^{S_2} &= \boldsymbol{r}_{_{S2}}(t_{_{S2}}) - \boldsymbol{r}_{_O}(t_{_O}).
\end{align}
The relative velocity and acceleration of the two satellites can then be expressed as
\begin{align}
\dot{\boldsymbol{r}}_{_O}^{S_i} &= \frac{d \boldsymbol{r}_{_O}^{S_i}}{dt} = \frac{d\boldsymbol{r}_{_{Si}}}{dt}(t_{_{Si}}) - \frac{d\boldsymbol{r}_{_O}}{dt}(t_{_O}),\\
\ddot{\boldsymbol{r}}_{_O}^{S_i} & = \frac{d^2\boldsymbol{r}_{_O}^{S_i}}{dt^2} = \frac{d^2\boldsymbol{r}_{_{Si}}}{dt^2}(t_{_{Si}}) - \frac{d^2\boldsymbol{r}_{_O}}{dt^2}(t_{_O}) ; i\in\{1,2\}. \label{eqn:inertialRelativeAcceleration}
\end{align}

As mentioned in Section \ref{sec:introduction}, a mutual approximation is defined as a point of closest encounter of two satellites in the field of view of an observer (see Figure \ref{fig:mutualApproxDefinition}). This corresponds to the moment at which the apparent distance between the two moons reaches a minimum. The apparent distance as seen by an observer is
\begin{align}
d = \sqrt{X^2 + Y^2},
\end{align} 
where $X$ and $Y$ are the coordinates of the relative position between the satellites, in the instrumental frame of the observer. 

The apparent relative position coordinates $X$ and $Y$ are defined as
\begin{align}
X &= (\alpha_{_{S2}} - \alpha_{_{S1}})\cos\left(\frac{\delta_{_{S1}} + \delta_{_{S2}}}{2}\right), \label{eqn:definitionX}\\
Y &= \delta_{_{S2}} - \delta_{_{S1}}. \label{eqn:definitionY}
\end{align}
$X$ and $Y$ thus depend on the right ascensions $\alpha_{_{Si}}$ and declinations $\delta_{_{Si}}$ of the two satellites, which are functions of the inertial relative range vectors with respect to the observer:
\begin{align}
\alpha_{_{S_{i}}}\left(\boldsymbol{r}_{_O}^{S_{i}}\right) &= 2 \arctan\left(\frac{y_{i}}{\sqrt{x_{i}^2+y_{i}^2}+x_{i}}\right), \label{eqn:definitionRightAscension}\\
\delta_{_{S_{i}}}\left(\boldsymbol{r}_{_O}^{S_{i}}\right) &=\frac{\pi}{2} - \arccos\left(\frac{z_{i}}{\sqrt{x_{i}^2+y_{i}^2+z_{i}^2}}\right), \label{eqn:definitionDeclination}
\end{align}
where $\left[x_{i},y_{i},z_{i}\right]$ correspond to the components of the relative range vectors $\boldsymbol{r}_{_O}^{S_{i}}$ (see Figure \ref{fig:referenceFrames}). $X$ and $Y$ are thus time-dependent, as they are indirectly defined by the time-varying relative range vectors between each of the two satellites and the observer. 
In the rest of this paper, $r_{i}$ denotes the norm of these relative range vectors and $r_{i_{xy}}$ the norm of the reduced vector $\left[x_{i},y_{i},0\right]$. $\delta_m$ refers to the average declination, such that 
$\delta_m = \left(\delta_{_{T_{1}}} + \delta_{_{T_{2}}} \right)/2$. The differences in right ascension and declination are noted $\Delta\alpha = \alpha_{_{T2}} - \alpha_{_{T1}}$  and $\Delta\delta = \delta_{_{T2}} - \delta_{_{T1}}$.

By definition, the central instant $t_c$ of a mutual approximation (recorded by the observer) fulfills the following condition: 
\begin{align}
\frac{d }{dt}\left(\sqrt{X(t_c)^2 + Y(t_c)^2}\right) = 0.
\end{align}
The apparent distance at $t_c$ is referred to as the impact parameter of the mutual approximation and denoted $d_c$.

\subsection{Analytical expressions for central instants} \label{sec:expressionCentralInstants}
The central instant $t_c$ is typically determined by fitting a fourth order polynomial to the apparent distance history between two satellites (see Figure \ref{fig:mutualApproxDefinition}). The roots of the derivative of the fitted polynomial provide the estimated central instant of the close encounter. For simulated mutual approximations, the procedure can be iterated to improve the precision of the predicted central instants by re-centring the polynomial fit on the current estimate of the point of closest approach.

A fourth order polynomial is needed to reproduce the relative motion of the two satellites over the typical duration of a close encounter (i.e. 60 minutes). However, when focusing on only a fraction of this event, a fourth order polynomial is not necessary. For instance, a second order polynomial provides a fit over the interval $\left[t_c-15\text{min} ; t_c\right.$ $\left.+15\text{min}\right]$  which is as good as the one provided by a fourth order polynomial over the whole event, as shown in Appendix \ref{appendix:polynomialFitting}.

To derive observation partials, we quantify the effect of very small changes in position and velocity of the two satellites. We are thus investigating only slight variations of the central instant $t_c$, and can limit our analysis to short time intervals centred on the current estimate of $t_c$. Consequently, for our analysis, it is safe to approximate the apparent relative motion of the two satellites by a second order polynomial only. 

Around the point of closest approach, the relative position coordinates $X$ and $Y$ can thus be expressed as a function of three polynomial coefficients each:
\begin{align}
X(t-t_c) &= a_0 + a_1 (t-t_c) + a_2(t-t_c)^2, \label{eqn:polynomialFunctionX}\\
Y(t-t_c) &= b_0 + b_1 (t-t_c) + b_2(t-t_c)^2. \label{eqn:polynomialFunctionY}
\end{align}
These polynomial coefficients are directly given by the apparent relative position, velocity, and acceleration coordinates at central instant $t_c$. Introducing the relative time $t^\prime = t-t_c$ as well as simplified notations ($X_c=X(t_c)$, $\dot{X}_c= \dot{X}(t_c)$, etc.), Equations \ref{eqn:polynomialFunctionX} and \ref{eqn:polynomialFunctionY} can be rewritten as follows:
\begin{align}
X(t^{\prime}) &= X_c + \dot{X}_c t^{\prime} + \frac{\ddot{X}_c}{2}{t^{\prime}}^2, \label{eqn:approxX}\\
Y(t^{\prime}) &= Y_c + \dot{Y}_c t^{\prime} + \frac{\ddot{Y}_c}{2}{t^{\prime}}^2. \label{eqn:approxY}
\end{align}
The relative velocity coordinates are then approximated by a first order polynomial when close enough to the central instant:
\begin{align} 
\dot{X}(t^{\prime}) &= \dot{X}_c + \ddot{X}_c t^{\prime}, \label{eqn:approxDotX}\\
\dot{Y}(t^{\prime}) &= \dot{Y}_c + \ddot{Y}_c t^{\prime}. \label{eqn:approxDotY}
\end{align} 

Higher-order terms could be included in Equations \ref{eqn:approxX}-\ref{eqn:approxDotY}. However, as discussed above, a second-order polynomial is well-suited to reproduce the apparent relative motion of the two satellites around the point of closest encounter. Higher-order terms can thus be safely neglected, as shown by the verification of our analytical partials for central instants (see Appendix \ref{appendix:validationPartials}).

As already mentioned in Section \ref{sec:mutualApproximationDefinition}, the derivative of the apparent distance is equal to zero at central instant $t_c$. Therefore, the dot product between the relative position and velocity vectors must be equal to zero, leading to the following condition:
\begin{align}
&\left(X_c + \dot{X}_c t^\prime + \frac{\ddot{X}_c}{2}{t^\prime}^2 \right)\left(\dot{X}_c + \ddot{X}_c t^\prime\right) \nonumber \\
&+ \left(Y_c + \dot{Y}_c t^\prime + \frac{\ddot{Y}_c}{2}{t^\prime}^2 \right)\left(\dot{Y}_c + \ddot{Y}_c t^\prime\right) = 0.
\end{align}
The above equation can be rewritten as a third-order polynomial expression in $t^{\prime}$:
\begin{align}
&\left(\ddot{X}_c^2 + \ddot{Y}_c^2\right){t^\prime}^3 + 3\left(\dot{X}_c \ddot{X}_c + \dot{Y}_c \ddot{Y}_c\right){t^\prime}^2 \nonumber \\
&+ 2\left(\dot{X}_c^2+ \dot{Y}_c^2 + X_c\ddot{X}_c + Y_c \ddot{Y}_c\right)t^\prime + 2\left(X_c \dot{X}_c + Y_c \dot{Y}_c\right) = 0. \label{eqn:cubicPolynomial}
\end{align}

Solving for $t^\prime$ is equivalent to finding the roots of this cubic polynomial, which can be done analytically with Cardano's formula \citep[e.g.][]{weisstein2002cubic}. In case the cubic polynomial has three real roots, the closest to the current $t_c$ estimate should be selected, the other two falling outside the nominal duration of the close encounter event in most cases. An analytical expression can thus be derived for $t^\prime$, as a function of the apparent position, velocity, and acceleration components at $t_c$:
\begin{align}
t^\prime &= f\left(X_c, Y_c, \dot{X}_c, \dot{Y}_c, \ddot{X}_c, \ddot{Y}_c\right). \label{eqn:functionCentralInstant}
\end{align}

Formulations for $\dot{X}$ and $\dot{Y}$ are derived from expressions for $X$ and $Y$ (Equations \ref{eqn:definitionX} and \ref{eqn:definitionY}), as follows:
\begin{align}
\dot{X} =& \Delta\dot{\alpha}\cos\left(\delta_m\right) - \Delta\alpha\sin\left(\delta_m\right)\dot{\delta}_m, \label{eqn:definitionVx} \\
\dot{Y} = &\Delta\dot{\delta}. \label{eqn:definitionVy}
\end{align}
$\dot{\alpha}$ and $\dot{\delta}$ can be computed from Equations \ref{eqn:definitionRightAscension} and \ref{eqn:definitionDeclination} as a function of the inertial relative position and velocity:
\begin{align}
\dot{\alpha}_{_{S_{i}}} &=\frac{x_{i} \dot{y}_{i} - y_{i} \dot{x}_{i}}{r_{i_{xy}}^2}, \label{eqn:firstDerivativeAlpha}\\
\dot{\delta}_{_{S_{i}}} &= \frac{-z_{i} \left(x_{i} \dot{x}_{i} + y_{i} \dot{y}_{i}\right) + r_{i_{xy}}^2\dot{z}_{i}}{r_{i}^2 r_{i_{xy}}} ; i \in {1,2}. \label{eqn:firstDerivativeDelta}
\end{align}

Finally, the apparent relative acceleration components $\ddot{X}$ and $\ddot{Y}$ are required and can be similarly derived:
\begin{align}
\ddot{X} &= \Delta\ddot{\alpha}\cos\left(\delta_m\right) - 2\dot{\delta}_m \Delta\dot{\alpha} \sin\left(\delta_m\right) \nonumber \\
& - \Delta\alpha \left({\dot{\delta}_m}^2 \cos\left(\delta_m\right) + \ddot{\delta}_m \sin\left(\delta_m\right) \right) \label{eqn:definitionAx} \\
\ddot{Y} & = \Delta\ddot{\delta}, \label{eqn:definitionAy}
\end{align}
where the second time derivatives of $\alpha$ and $\delta$ also depend on the inertial relative acceleration:
\begin{align}
\ddot{\alpha}_{_{S_{i}}} &= \frac{-2\left(x_{i} \dot{y}_{i} - y_{i} \dot{x}_{i}\right)\left(x_{i} \dot{x}_{i} + y_{i} \dot{y}_{i}\right)}{r_{i_{xy}}^4} + \frac{\left(x_{i} \ddot{y}_{i} - y_{i} \ddot{x}_{i}\right)}{r_{i_{xy}}^2}, \label{eqn:secondDerivativeAlpha} \\
\ddot{\delta}_{_{S_{i}}} &= \frac{1}{r_{i}^2 r_{i_{xy}}} \left[-z_{i} \left(x_{i}\ddot{x}_{i} + y_{i} \ddot{y}_{i}\right) + r_{i_{xy}}^2\ddot{z}_{i} \right. \nonumber \\
&\left.-z_{i} \frac{\left(x_{i} \dot{y}_{i} - y_{i} \dot{x}_{i}\right)^2}{r_{i_{xy}}^2}\right. \nonumber  \\
&\left.+\frac{2 \left(\boldsymbol{r}_{_O}^{S_{i}} \cdot \dot{\boldsymbol{r}}_{_O}^{S_{i}}\right)}{r_{i}^2} \left(z_{i}(x_{i} \dot{x}_{i} + y_{i} \dot{y}_{i})- \dot{z}_{i}r_{i_{xy}}^2\right)\right] ; i\in\{1,2\}. \label{eqn:secondDerivativeDelta}
\end{align}
Inserting Equations \ref{eqn:definitionRightAscension}-\ref{eqn:definitionDeclination}, \ref{eqn:firstDerivativeAlpha}-\ref{eqn:firstDerivativeDelta}, and \ref{eqn:secondDerivativeAlpha}-\ref{eqn:secondDerivativeDelta} into Equations \ref{eqn:definitionX}-\ref{eqn:definitionY}, \ref{eqn:definitionVx}-\ref{eqn:definitionVy}, and \ref{eqn:definitionAx}-\ref{eqn:definitionAy} gives a direct analytical expression for $t^{\prime}$, and therefore for the central instant $t_c$, via Equation \ref{eqn:functionCentralInstant}.

\subsection{Partials with respect to the natural satellites' states} \label{sec:centralInstantPartials}

To estimate ephemerides using central instants as observables, the partials of $t_c$ with respect to the states of the two natural satellites are required. Recalling the analytical expression obtained for $t^\prime$ (Equation \ref{eqn:functionCentralInstant}) and noting $\boldsymbol{q}$ the vector of parameters, the central instants partials are 
\begin{align}
\frac{\partial t^\prime}{\partial \boldsymbol{q}} = &\frac{\partial f\left(X_c, Y_c, \dot{X}_c, \dot{Y}_c, \ddot{X}_c,\ddot{Y}_c\right)}{\partial \boldsymbol{q}} \\
= &\frac{\partial f}{\partial [X_c,Y_c]}\frac{\partial [ X_c,Y_c]}{\partial \boldsymbol{q}} + \frac{\partial f}{\partial [\dot{X}_c,\dot{Y}_c]}\frac{\partial [\dot{X}_c,\dot{Y}_c]}{\partial \boldsymbol{q}} \nonumber \\
&+ \frac{\partial f}{\partial [ \ddot{X}_c,\ddot{Y}_c]}\frac{\partial [\ddot{X}_c,\ddot{Y}_c]}{\partial \boldsymbol{q}} \label{eqn:partials_tc}.
\end{align} 

The partials of the relative apparent position, velocity, and acceleration can be decomposed as a function of the partials of $\alpha_{_{S_{i}}}$, $\delta_{_{S_{i}}}$, $\dot{\alpha}_{_{S_{i}}}$, $\dot{\delta}_{_{S_{i}}}$, $\ddot{\alpha}_{_{S_{i}}}$, and $\ddot{\delta}_{_{S_{i}}}$, as follows:
\begin{align}
\frac{\partial \left[X,Y\right]}{\partial \boldsymbol{q}} &= \frac{\partial[X,Y]}{\partial \left[\alpha,\delta\right]_{_{S_{i}}}}\frac{\partial \left[ \alpha, \delta \right]_{_{S_{i}}}}{\partial \boldsymbol{q}}, \\
\frac{\partial [\dot{X},\dot{Y}]}{\partial \boldsymbol{q}} &= \frac{\partial[\dot{X},\dot{Y}]}{\partial \left[\alpha,\delta\right]_{_{S_{i}}}}\frac{\partial \left[\alpha,\delta\right]_{_{S_{i}}}}{\partial \boldsymbol{q}} +\frac{\partial[\dot{X},\dot{Y}]}{\partial [\dot{\alpha},\dot{\delta}]_{_{S_{i}}} }\frac{\partial[\dot{\alpha},\dot{\delta}]_{_{S_{i}}} }{\partial \boldsymbol{q}}, \\
\frac{\partial \left[\ddot{X},\ddot{Y}\right]}{\partial \boldsymbol{q}} &= \frac{\partial[\ddot{X},\ddot{Y}]}{\partial [\alpha,\delta ]_{_{S_{i}}}}\frac{\partial [\alpha,\delta ]_{_{S_{i}}}}{\partial \boldsymbol{q}}+\frac{\partial[\ddot{X},\ddot{Y}]}{\partial [ \dot{\alpha},\dot{\delta}]_{_{S_{i}}}}\frac{\partial [ \dot{\alpha},\dot{\delta} ]_{_{S_{i}}} }{\partial \boldsymbol{q}} \nonumber \\
&+\frac{\partial[\ddot{X},\ddot{Y}]}{\partial [ \ddot{\alpha},\ddot{\delta} ]_{_{S_{i}}}}\frac{\partial [ \ddot{\alpha},\ddot{\delta} ]_{_{S_{i}}} }{\partial \boldsymbol{q}} ; i\in\{1,2\}.
\end{align}
From the definition of the apparent position ($X$,$Y$), velocity ($\dot{X}$,$\dot{Y}$), and accelerations ($\ddot{X}$,$\ddot{Y}$) in Equations \ref{eqn:definitionX}-\ref{eqn:definitionY}, \ref{eqn:definitionVx}-\ref{eqn:definitionVy}, and \ref{eqn:definitionAx}-\ref{eqn:definitionAy}, their partials with respect to the satellites' states can be easily derived (the proof is left as an exercise to the reader). Finally, the partials of $\alpha$, $\delta$, $\dot{\alpha}$, $\dot{\delta}$, $\ddot{\alpha}$, and $\ddot{\delta}$ also need to be computed with respect to the position and velocity vectors of the two satellites. 

To quantify the influence of the uncertainties in the observer's state on the estimated solution, partials with respect to $\boldsymbol{r}_{_O}$ and $\dot{\boldsymbol{r}}_{_O}$ might also be required. All derivations are provided in Appendix \ref{appendix:partials}. Our analytical formulation for the partials of the central instants with respect to both the satellites' and observer's states were validated numerically. The results of this verification are reported in Appendix \ref{appendix:validationPartials}.

\subsection{Light-time effects} \label{sec:lightTimeCorrections}
In Section \ref{sec:centralInstantPartials}, the contribution of the light-time effects was not yet included in the observation partials and we therefore assumed that both $t_{_{O}}$ and $t_{_{Si}}$ were fixed. Corrections required to account for the finite speed of light are now discussed.

When computing light-time effects, we typically fix either the time at the observed body (here $t_{_{Si}}$) or the time at the observer ($t_{_O}$). The other one is determined via an iterative scheme to ensure that the difference between the two times matches the light-time calculated from the observer and observed bodies' states \citep{moyer2000formulation}. For mutual approximations, the reception time should always be fixed. Fixing the two transmission times would indeed lead to two different inconsistent reception times for a unique observation. The light-time equations are expressed as follows \citep{moyer2000formulation}:
\begin{align}
t_{_{Si}} - t_{_{O}} &= \frac{\left|\boldsymbol{r}_{_{Si}}(t_{_{Si}}) - \boldsymbol{r}_{_O}(t_{_O})\right|}{c} ; i\in \{1,2\}, \label{eqn:LTtransmitter}
\end{align}
where $c$ refers to the speed of light and the observation time $t_{O}$ is a fixed unique value.

The partials of the light-time with respect to a vector of parameters $\boldsymbol{q}$ can then be derived from Equation \ref{eqn:LTtransmitter}: 
\begin{align}
\frac{\partial t_{_{Si}}}{\partial \boldsymbol{q}} &=\frac{1}{c} \frac{\boldsymbol{r}_{_O}^{S_i}}{r_{_{O}}^{Si}}\left( \frac{\partial\boldsymbol{r}_{_{Si}}}{\partial\boldsymbol{q}}(t_{_{Si}}) -  \frac{\partial\boldsymbol{r}_{_{O}}}{\partial \boldsymbol{q} }(t_{_{O}})+ \dot{\boldsymbol{r}}_{_{Si}}(t_{_{Si}})\frac{\partial t_{_{S1}}}{\partial \boldsymbol{q}}\right).
\end{align}
Solving for the partials of $t_{_{Si}}$ with respect to $\boldsymbol{q}$, we finally obtain \citep{moyer2000formulation}
\begin{align}
\frac{\partial t_{_{Si}}}{\partial \boldsymbol{q}} &= \frac{1}{c}\frac{ \boldsymbol{r}_{_{O}}^{Si}}{  r_{_{O}}^{Si}-\boldsymbol{r}_{_{O}}^{Si}\cdot \frac{\dot{\boldsymbol{r}}_{_{Si}}}{c}}\left(\frac{\partial\boldsymbol{r}_{_{Si}}}{\partial\boldsymbol{q}}(t_{_{Si}}) -  \frac{\partial\boldsymbol{r}_{_{O}}}{\partial \boldsymbol{q} }(t_{_{O}})\right).
\end{align}

The time $t_{_{Si}}$ thus depends on both the natural satellite's and observer's states. As already mentioned, right ascension and declination partials with respect to the vector of parameters $\boldsymbol{q}$ were provided for fixed $t_{_{O}}$ and $t_{_{Si}}$ in Section \ref{sec:centralInstantPartials}. When accounting for the light-time effect, the complete formulation for those partials becomes
\begin{align}
\frac{\partial \left[\alpha,\delta\right]_{_{S_{i}}}}{\partial \boldsymbol{q}} &= \left.\frac{\partial \left[\alpha,\delta\right]_{_{S_{i}}}}{\partial \boldsymbol{q}}\right|_{t_{_{Si}}}+\left[\dot{\alpha},\dot{\delta}\right]_{_{S_{i}}}\frac{\partial t_{_{Si}}}{\partial \boldsymbol{q}} ; \text{ } i\in \{1,2\}.
\end{align}
The same applies to the partials of $\dot{\alpha}$, $\dot{\delta}$, $\ddot{\alpha}$, and $\ddot{\delta}$, and leads to the following expressions:
\begin{align}
\frac{\partial [\dot{\alpha},\dot{\delta}]_{_{S_{i}}}}{\partial \boldsymbol{q}} &= \left.\frac{\partial [\dot{\alpha},\dot{\delta}]_{_{S_{i}}}}{\partial \boldsymbol{q}}\right|_{t_{_{Si}}}+[\ddot{\alpha},\ddot{\delta}]_{_{S_{i}}}\frac{\partial t_{_{Si}}}{\partial \boldsymbol{q}}, \\
\frac{\partial [\ddot{\alpha},\ddot{\delta}]_{_{S_{i}}}}{\partial \boldsymbol{q}} &= \left.\frac{\partial [\ddot{\alpha},\ddot{\delta}]_{_{S_{i}}}}{\partial \boldsymbol{q}}\right|_{t_{_{Si}}}+[\dddot{\alpha},\dddot{\delta}]_{_{S_{i}}}\frac{\partial t_{_{Si}}}{\partial \boldsymbol{q}} ; \text{ } i\in \{1,2\}. \label{eqn:LTpartialsSecondTimeDerivativeAlphaDelta}
\end{align}

According to Equation \ref{eqn:LTpartialsSecondTimeDerivativeAlphaDelta}, the complete partials for $\ddot{\alpha}_{_{S_{i}}}$ and $\ddot{\delta}_{_{S_{i}}}$ require one to compute $\dddot{\alpha}_{_{S_{i}}}$ and $\dddot{\delta}_{_{S_{i}}}$ (see Equation \ref{eqn:LTpartialsSecondTimeDerivativeAlphaDelta}), and thus the time derivative of the relative acceleration of each satellite with respect to the observer. This would significantly increase both the implementation and computational efforts, while the $\ddot{\alpha}_{_{S_{i}}}$ and $\ddot{\delta}_{_{S_{i}}}$ partials only marginally contribute to the central instant partials (at most of the order of $0.001\%$ for the case of the Galilean satellites, see Appendix \ref{appendix:LTcontributions}, Table \ref{tab:contributionPartialsSecondTimeDerivativeRightAscensionDeclination}). 
As a consequence, neglecting the light-time effects when computing the partials for $\ddot{\alpha}_{_{S_{i}}}$ and $\ddot{\delta}_{_{S_{i}}}$ is a fair simplifying assumption, which was applied in the rest of this study.  

\subsection{Alternative observables} \label{sec:alternativeObservables}
As already mentioned in Section \ref{sec:introduction},  the alternative observable recommended by \cite{morgado2019approx} correspond to the derivative of the apparent distance, defined as
\begin{align}
h &= \frac{d }{dt}\left(\sqrt{X^2 + Y^2}\right)=\frac{X \dot{X}+Y\dot{Y}}{\sqrt{X^2 + Y^2}}. \label{eqn:alternativeObservable}
\end{align}
If $X$ and $Y$ and their time derivatives $\dot{X}$ and $\dot{Y}$ were computed at the exact central instant $t_c$ of the close encounter, the observable $h$ would by definition be equal to zero. This is however not the case. This observable thus indirectly evaluates the difference between the current estimate of $t_c$ and its true value by quantifying how much the derivative of the apparent distance departs from zero.

In contrast to central instants which also depend on the satellites' relative accelerations, alternative observables are thus only a function of their relative position and velocity. The partials of such an observable with respect to a vector of parameters $\boldsymbol{q}$ are much easier to derive than for central instants and are given by \cite{morgado2019approx}:
\begin{align}
\frac{\partial h}{\partial \boldsymbol{q}} =& \frac{1}{\sqrt{X^2 + Y^2}}\left(X \frac{\partial \dot{X}}{\partial \boldsymbol{q}} + \dot{X}\frac{\partial X}{\partial \boldsymbol{q}}+ Y \frac{\partial \dot{Y}}{\partial \boldsymbol{q}}+ \dot{Y}\frac{\partial Y}{\partial \boldsymbol{q}}\right) \nonumber \\
&-\frac{X\dot{X}+Y\dot{Y}}{\left(X^2+Y^2\right)^{3/2}}\left(X \frac{\partial X}{\partial \boldsymbol{q}}+ Y\frac{\partial Y}{\partial \boldsymbol{q}}\right).
\end{align}
The results of the comparison between the two types of observables are discussed in Section \ref{sec:results}.

\section{Observations simulation and ephemerides estimation} \label{sec:stateEstimationProcedure}

We first describe how mutual approximations are simulated in Section \ref{sec:mutualApproxSimulation}, before introducing the covariance analysis used to compare the two observable types in Section \ref{sec:covarianceAnalysis}. The strategy applied to weight the mutual approximations' observables is then discussed in Section \ref{section:dataWeights}. Finally, Section \ref{sec:contributionsDefinition} defines an additional figure of merit to analyse the estimation solution.

\subsection{Mutual approximations simulation} \label{sec:mutualApproxSimulation}
We used simulated mutual approximations in our analysis. As a preliminary test case, we first propagated the trajectories of Io and Europa only and detected close encounters between these two moons (results discussed in Sections \ref{sec:directComparison20202029} to \ref{sec:influenceWeightingScheme}). A more complete simulation including all Galilean moons was also conducted to verify the findings of the former simple test case (Section \ref{sec:validationFourGalileanMoons}).  

The orbits of the Galilean moons were propagated using a simplified dynamical model. For each of the moons, we considered only the point-mass gravitational accelerations exerted by Jupiter and the three other satellites. A more detailed dynamical model \citep[e.g.][]{dirkx2016dynamical,lainey2004new} would yield more accurate propagated orbits for the Galilean moons, and thus affect the predicted mutual approximations. However, we focus on comparing two types of mutual approximations' observables. High-accuracy dynamical modelling is therefore not critical for this study, as long as the same set of simulated observations is used for both observable types.

Mutual approximations were simulated for the period 2020-2029. To limit the number of observations, we only considered mutual approximations with an impact parameter lower than 30 arcseconds (as), in accordance with \cite{morgado2019approx}. We selected three of the ground stations involved in the 2016-2018 observational campaign reported in \cite{morgado2019approx}, designated by FOZ, OHP, and OPD (their coordinates are reported in Table \ref{tab:groundStationsCoordinates}). 
To ensure the feasibility of the observation, events which would be observable during daytime were discarded. In addition, the lower limit on the distance between the mutual approximation and the limb of Jupiter was set to 10 as. Only mutual approximations that would be observed from the three ground stations under an elevation angle larger than 30 degrees were included. 

\begin{table}[h]
	\caption{Ground stations' geodetic coordinates.}
	\label{tab:groundStationsCoordinates}
	\centering
	\resizebox{1.0\columnwidth}{!}{
		\begin{tabular}{l l l c}
			\hline \hline \noalign{\vskip 0.02in}
			Alias &  &  &  \\ 
			Site & Longitude [E] & Latitude [N] & Altitude [m] \\
			Location & & \\ \hline \noalign{\vskip 0.02in}
			FOZ & & &  \\
			Foz do Iguacu & - 54$^{\circ}$35'37.0'' & - 25$^{\circ}$26'05.0''  & 184 \\ 
			PR-Brazil & & & \\ \hline \noalign{\vskip 0.02in}
			OHP & & & \\ 
			Haute-Provence & 05$^{\circ}$42'56.5'' & 43$^{\circ}$55'54.7'' & 633  \\ 
			France & & & \\ \hline \noalign{\vskip 0.02in}
			OPD & & & \\
			Itajuba & - 45$^{\circ}$34'57.5'' W & - 22$^{\circ}$32'07.8'' & 1864  \\
			MG-Brazil & & & \\
			\hline                     
	\end{tabular}}
	\tablefoot{The three ground stations reported in this table are the ones from which the observations of the mutual approximations are simulated. The table is adapted from \cite{morgado2019approx}.}
\end{table}

When achievable under the aforementioned conditions, a single event can be observed by several ground stations. Those multiple observations of one mutual approximation were assumed to have uncorrelated noise and thus they were added as independent observations to the state estimation. This implies that such simultaneous observations improve the estimation solution by increasing the size of the observational data set, as formal errors are expected to scale down with $\sqrt{n}$ ($n$ being the total number of observations).

Finally, weather conditions were taken into account to obtain a realistic set of observations. Due to bad weather conditions, about 35\% of the predicted mutual approximations could not be observed during the 2016-2018 campaign \citep{morgado2019approx}. We took a conservative approach to simulate these bad weather conditions and discarded 50\% of the viable observations, selected arbitrarily using a uniform distribution.

\begin{figure*} 
	\centering
	\begin{subfigure}{0.49\textwidth}
		\centering
		{\includegraphics[width=1.0\textwidth]{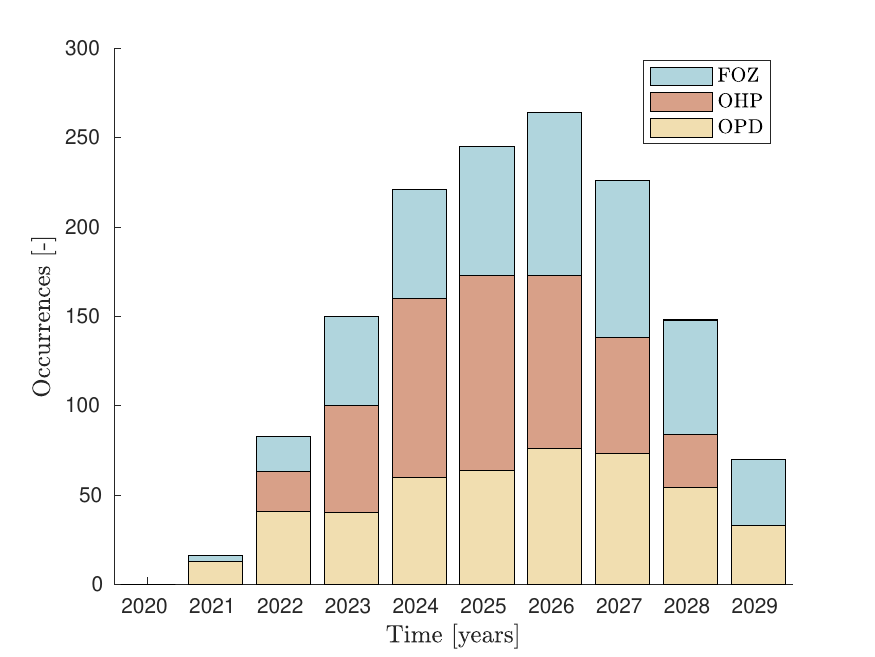}}
		\caption{ }
		\label{fig:observationsPerYearPerGroundStation}
	\end{subfigure}
	\hfill{}
	\begin{subfigure}{0.49\textwidth}
		\centering
		{\includegraphics[width=1.0\textwidth]{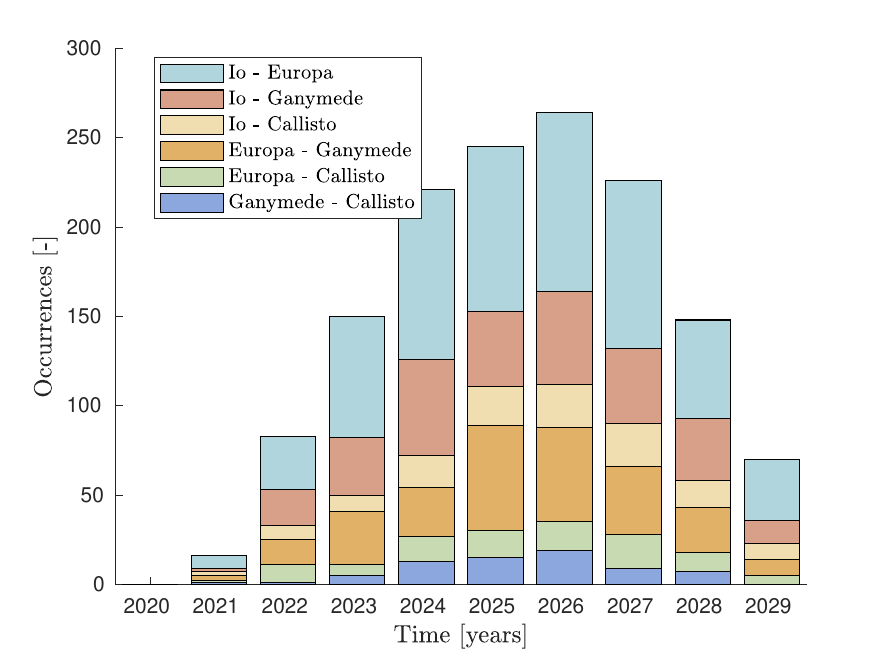}}
		\caption{ }
		\label{fig:observationsPerYearPerMoons}
	\end{subfigure}
	\caption{Distribution of the simulated mutual approximations per year, depending on the ground station (panel a) and on the two moons involved (panel b). Mutual approximations which have been discarded to mimic the effect of bad weather conditions are not included in this distribution.}
	\label{fig:observationDistribution}
\end{figure*}

The distribution per year of the remaining simulated mutual approximations is shown in Figure \ref{fig:observationDistribution}. Figure \ref{fig:observationsPerYearPerGroundStation} displays the fraction of simulated events per ground station, while Figure \ref{fig:observationsPerYearPerMoons} focuses on the number of mutual approximations for each combination of two Galilean moons. It is interesting to note that no mutual approximation respecting the conditions mentioned in the previous paragraphs could be found in 2020, and that some years are more favourable to such events, due to the time evolution of the  Earth - Jovian system relative geometry.

\subsection{Covariance analysis} \label{sec:covarianceAnalysis}
To compare state estimations obtained with the two types of mutual approximation observables, we limited ourselves to a covariance analysis. Despite its  limitations (Gaussian observation noise, dynamical and observational models assumed perfect), such an analysis is well-adapted for comparison purposes. Formal errors are known to be too optimistic compared to true errors, but we only focus on comparing two sets of estimation errors and not on absolute error values. Since mutual approximations are almost exclusively sensitive to the relative dynamics between the two satellites while both their absolute states are estimated, realistic errors are anyway difficult to achieve without including other observations.

In our simulations, the estimated parameters were the initial states of the moons involved in the mutual approximations. In most of our analysis, only the Jupiter-centred initial states of Io and Europa are estimated (Sections \ref{sec:directComparison20202029} to \ref{sec:influenceWeightingScheme}), while we also solved for the initial states of Ganymede and Callisto in the more complete case used for verification (see Section \ref{sec:validationFourGalileanMoons}). For the moons' initial position components, a priori covariance of 100 km was considered, while it was set to 100 m/s for their initial velocity. These a priori values are large, but were only included to slightly constrain the estimation, thus avoiding an ill-posed problem and making the comparison between the two observable types' estimation solutions possible. 

\subsection{Data weights} \label{section:dataWeights}

Observation weights are usually applied to account for the quality of the data. For our comparative analysis, it is essential to ensure that the data weights are consistent between the two types of observables. 
We	used an error of 3.5 s for the central instants $t_c$ \citep[average error obtained over the 104 observed mutual approximations of the 2016-2018 campaign reported in][]{morgado2019approx}.

To derive appropriate weights for the alternative observables, the shape of the simulated mutual approximation must be taken into account. By definition, the derivative of the apparent distance (i.e. alternative observable) is always equal to zero at $t = t_c$. However, an error of 3.5 seconds in the determination of the central instant would shift this value away from zero. The exact value of the resulting alternative observable error directly depends on the specific geometry of each mutual approximation. The alternative observable error was thus individually computed for each observation, as follows:
\begin{align}
\sigma_{\mathrm{alt.}} = \frac{\left|\dot{d}(t_c-\sigma_{t_c})\right| + \left| \dot{d}(t_c+\sigma_{t_c})\right|}{2}, \label{eqn:alternativeObservableWeight}
\end{align}
where $\sigma_{t_c}$ is set to its averaged value ($\sigma_{t_c}$ = 3.5 s) and $\dot{d}$ is the derivative of the apparent distance (given by Equation \ref{eqn:alternativeObservable}).

Consistent weights between the two observables are not only needed to perform a meaningful comparison. When using alternative observables, weighting can be an indirect way to account for the satellites' relative dynamics during the close encounter. Indeed, a non-zero value for the derivative of the apparent distance at $t_c$ only quantifies how much the observed central instant departs from the current point of closest approach. However, it does not provide any information about the current apparent distance minimum. For a given non-zero value of the apparent distance derivative, the difference between the observed and current central instants entirely depends on the satellites' apparent relative dynamics, which drive the geometry of the observed encounter.

This effect is, by definition, inherently captured by central instants, for which applying an appropriate constant weight value is thus suitable. For alternative observables, on the other hand, individual weights accounting for each mutual approximation's dynamics, as given in Equation \ref{eqn:alternativeObservableWeight}, are crucial. This is necessary to translate an error in the estimated central instant to an error in the derivative of the apparent distance.

The importance of applying this weighting strategy to obtain consistent estimation solutions with the two different observable types is demonstrated in Section \ref{sec:influenceWeightingScheme}. Furthermore, we computed the appropriate alternative observables' weights for the past mutual approximations observed during the 2016-2018 campaign and reported in \cite{morgado2019approx}. These weights are provided in Appendix \ref{appendix:weightsPastObservations} and should be used when including the 2016-2018 mutual approximations in the state estimation.

\subsection{Contribution of each observation to the solution} \label{sec:contributionsDefinition}
To perform a detailed comparison of the two observable types, mutual approximations' contributions to the solution were used as an additional figure of merit to complement the covariance analysis.
In this study, each observation's contribution to the solution is defined as the root-mean-square (RMS) of the weighted observation partial with respect to the parameters of interest's vector $\boldsymbol{q}$. 

For example, the contribution $c$ of an observation $h$ to Io's Jupiter-centred initial position vector is expressed as:
\begin{align}
c_{\left(\boldsymbol{r}_{_{\mathrm{Io}}}\right)}(h) = \sqrt{\left(\frac{\partial h}{\partial x_{_{\mathrm{Io}}}(t_0)}\right)^2 + \left(\frac{\partial h}{\partial y_{_{\mathrm{Io}}}(t_0)}\right)^2 + \left(\frac{\partial h}{\partial z_{_{\mathrm{Io}}}(t_0)}\right)^2},
\end{align}
where $t_0$ is the initial epoch at which Io's state is estimated. The contribution $c_{\left(\boldsymbol{q}\right)}(h)$ to the vector of parameters $\boldsymbol{q}$ is then normalised as follows (the bar indicates normalisation):
\begin{align}
\bar{c}_{\left(\boldsymbol{q}\right)}(h) = \frac{\log\left(c_{\left(\boldsymbol{q}\right)}(h)\right) - \log\left(\min(\boldsymbol{c}_{\left(\boldsymbol{q}\right)})\right)}{\log\left(\max\left(\boldsymbol{c}_{\left(\boldsymbol{q}\right)}\right)\right) - \log\left(\min\boldsymbol{c}_{\left(\boldsymbol{q}\right)})\right)}, \label{eqn:normalisedContribution}
\end{align}
where $\boldsymbol{c}_{\left(\boldsymbol{q}\right)}$ is the vector containing the contributions of the entire set of mutual approximations with respect to $\boldsymbol{q}$ (for one type of observable).

\section{Results} \label{sec:results}
We present here the results of the comparison between the ephemeris estimation determination solutions obtained using central instants and alternative observables, respectively. The comparison is first conducted for a simple test case with Io and Europa only, to analyse how each mutual approximation contributes to the ephemerides solution and how this affects the relative performance of the two types of observables. Results of this first analysis are presented in Sections \ref{sec:directComparison20202029} to \ref{sec:influenceWeightingScheme}. A more complete test case also including Ganymede and Callisto is used to verify those findings (Section \ref{sec:validationFourGalileanMoons}). 

\subsection{Comparison over the 2020-2029 observational period} \label{sec:directComparison20202029}

\begin{figure*}
	\centering
	\includegraphics[width=17cm]{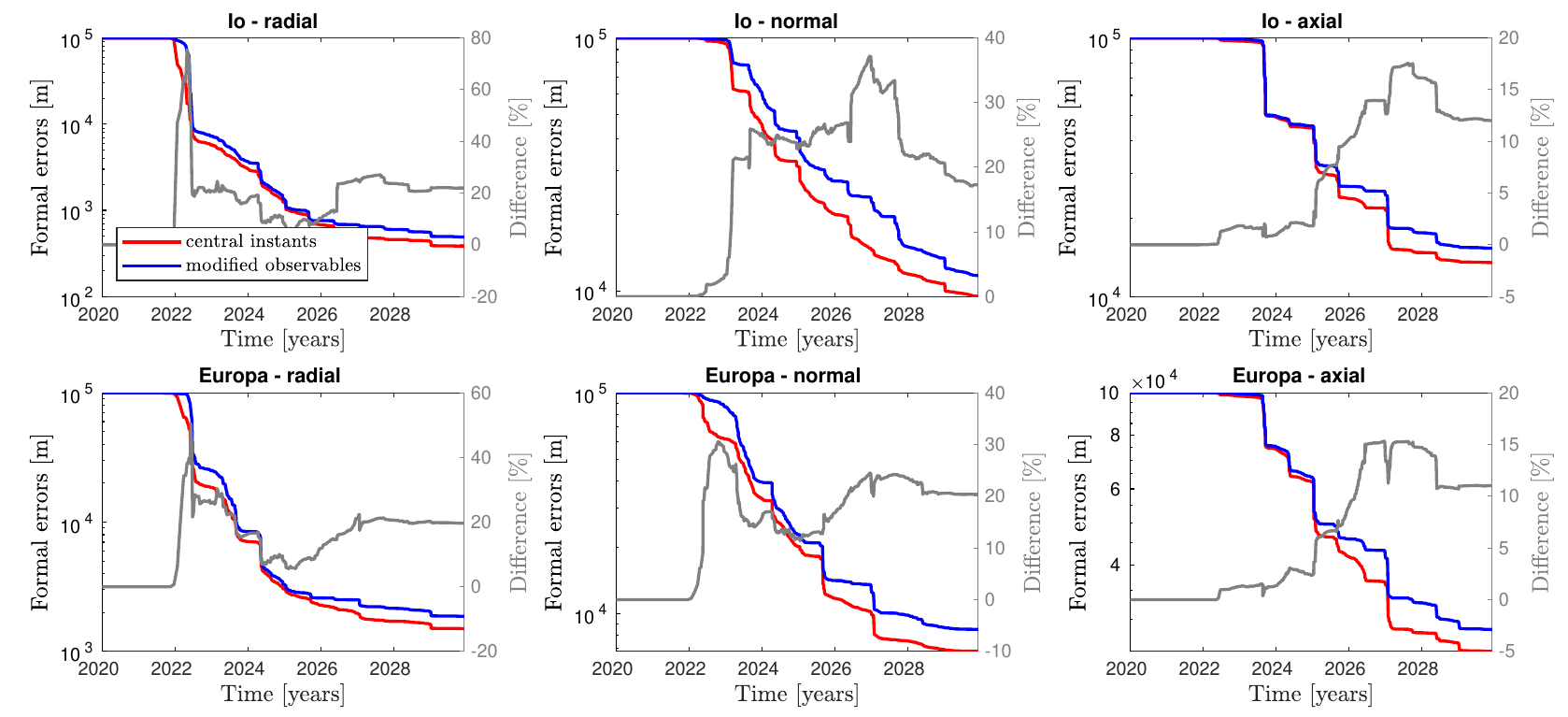}
	\caption{Time evolution of the formal errors in Io's and Europa's initial RSW coordinates (radial, normal, and axial directions, see Figure \ref{fig:referenceFrames}), as more observations are progressively included in the state estimation. The blue line displays the time evolution of the formal errors obtained by using the central instants as observables (left y-axis). The red line corresponds to the time evolution of the formal errors obtained from alternative observables (left y-axis too). The initial values of the formal errors, before including any observation, correspond to the a priori covariance values used for the regularisation (i.e. 100 km, see Section \ref{sec:covarianceAnalysis}). The grey line (right y-axis) represents the relative difference (in percentage) between the two solutions as a function of time. The formal errors are equal to their initial values until the inclusion of the first mutual approximation (towards the end of 2021) and no difference between the two observables' solutions is thus observed beforehand.}
	\label{fig:formalErrorsHistoryIoEuropaOnly}
\end{figure*}

We first only simulated mutual approximations between the two innermost Galilean moons, for the period 2020 - 2029, and estimated Io's and Europa's initial states from those simulated observations. The evolution of the formal errors with time is displayed in Figure \ref{fig:formalErrorsHistoryIoEuropaOnly}, as more mutual approximations are included in the estimation. The differences in formal errors between the two types of observables do not exceed 20\% at the end of our simulation, after ten years of observations. Alternative observables and central instants lead to comparable formal errors evolutions. At first order, this proves that the two types of mutual approximations' observables are largely equivalent, at least when enough observations are added to the state estimation. It validates the recommendations formulated in \cite{morgado2019approx}, but seems to contradict the results on numerical partials in \cite{emelyanov2017precision}.

Nonetheless, using central instants still results in slightly lower formal errors for each component of both Io's and Europa's initial position. The formal error improvement is stronger in the radial and normal directions (about 20\% for both Io and Europa at the end of simulation) and less significant in the axial direction (only 10-12\%). As mentioned in Section \ref{sec:alternativeObservables}, the observation partials developed for the central instants account for variations in the apparent relative acceleration between the two satellites, while this is not the case for alternative observables. The additional information captured by central instants thus principally lies within the orbital plane of the Galilean moons, within which the inter-moons accelerations primarily lie. On the other hand, the central instants are not significantly more sensitive than alternative observables to state variations in the axial direction.

Interestingly, the difference in formal errors between the two types of observables is not constant over time, as clearly highlighted by Figure \ref{fig:formalErrorsHistoryIoEuropaOnly}. It can be as low as a few percents (e.g. Io's normal position in year 2021) or as high as 35\% (e.g. Io's normal position during the first half of year 2027). This is related to the mutual approximations' heterogeneous contribution to the solution: it varies from one observation to another, but also between the two observable types. The cause of this heterogeneity is further discussed in Section \ref{sec:influenceMutualApproxCharac}.

\begin{figure*} 
	\centering
	\begin{subfigure}{0.49\textwidth}
		\centering
		{\includegraphics[width=1.0\textwidth]{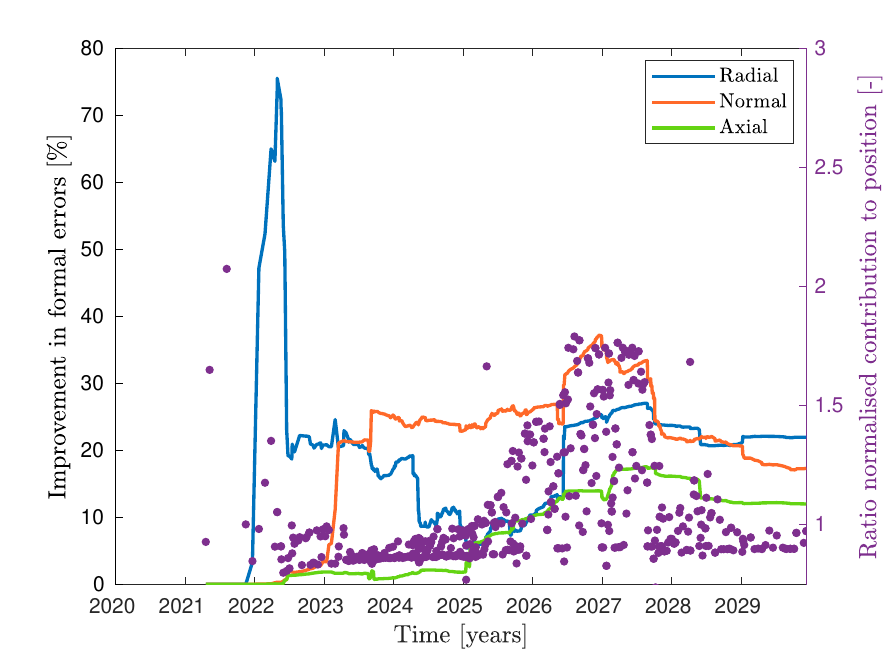}}
		\caption{} 
		\label{fig:improvementVsContributionIo}
	\end{subfigure}
	\hfill{} 
	\begin{subfigure}{0.49\textwidth}
		\centering
		{\includegraphics[width=1.0\textwidth]{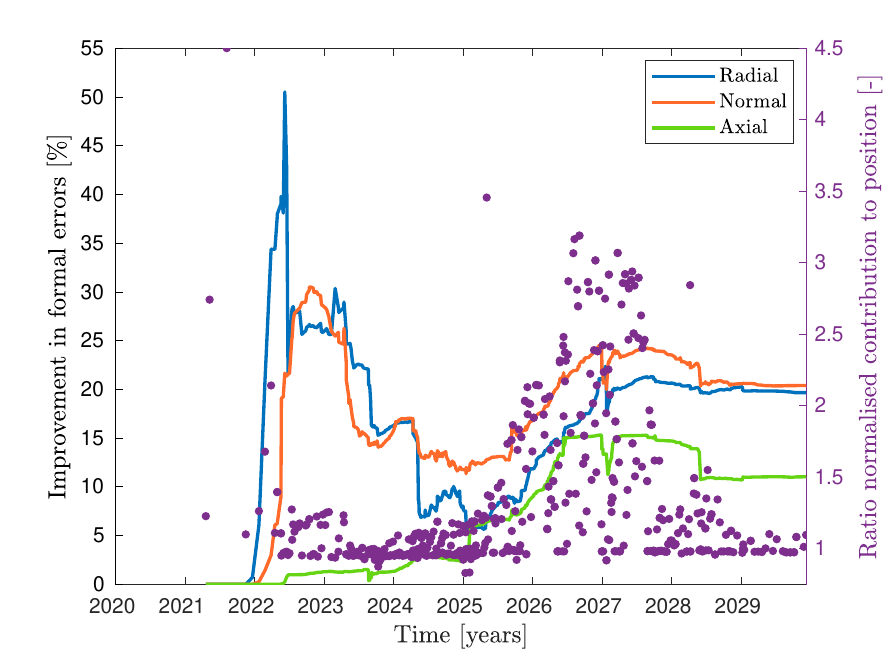}}
		\caption{}
		\label{fig:improvementVsContributionEuropa}
	\end{subfigure}
	\caption{Reduction in formal errors obtained by using central instants instead of alternative observables, as more observations are added to the solution. It is displayed on the left axis, for the three position components in RSW frame (panel a: Io, panel b: Europa). On the right axis (purple dots), the ratio between the normalised contributions of central instants over their corresponding alternative observables is displayed, for each mutual approximation (normalised contributions are computed as in Equation \ref{eqn:normalisedContribution}). As mentioned in Section \ref{sec:mutualApproxSimulation}, the first simulated observation only occurs towards the end of 2021, hence the lack of data beforehand.}
	\label{fig:improvementVsContribution}
\end{figure*}

\begin{figure*}
	\centering
	\begin{subfigure}{0.49\textwidth}
		\centering
		{\includegraphics[width=1.0\textwidth]{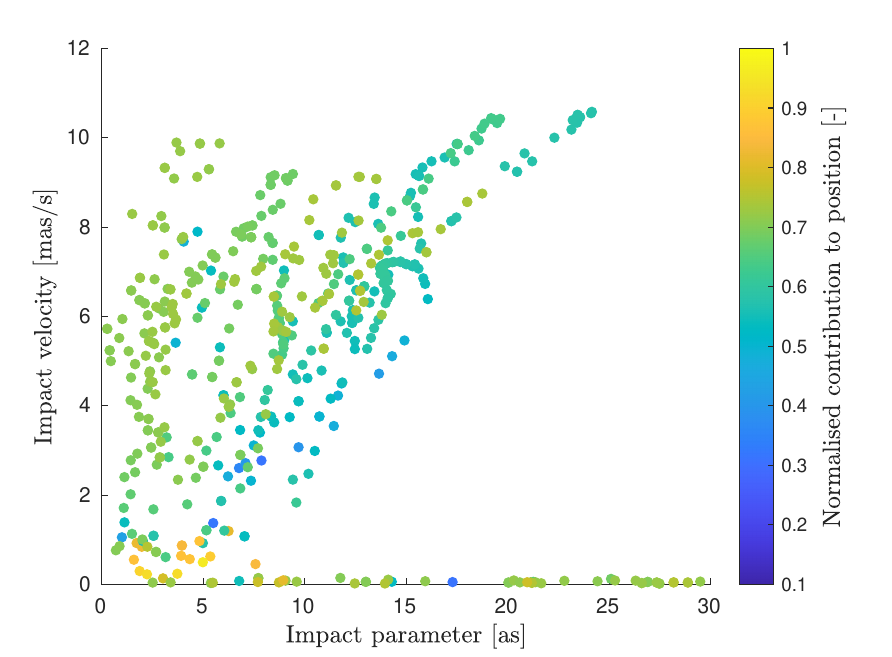}}
		\caption{} 
		\label{fig:parametersContributions_centralInstants}
	\end{subfigure}
	\begin{subfigure}{0.49\textwidth}
		\centering
		{\includegraphics[width=1.0\textwidth]{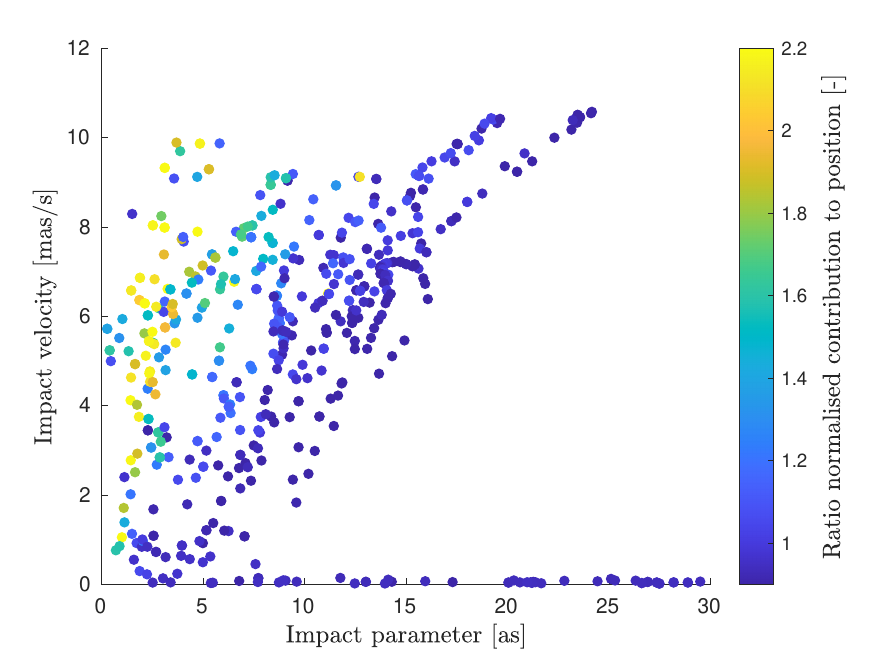}}
		\caption{} 
		\label{fig:ratioContributionsGeometryAveraged}
	\end{subfigure}
	\caption{Observations' contributions to the estimated initial positions' solution (in colours), as a function of each mutual approximation's characteristics (impact parameters and impact velocities, reported on the x- and y-axes, respectively). Panel a: normalised contribution of each mutual approximation to the solution (averaged between Io and Europa), when using central instants. Panel b: ratio between the normalised contributions of central instants over their corresponding alternative observables. The normalised contribution of each mutual approximation to the solution is obtained with Equation \ref{eqn:normalisedContribution}.}
	\label{fig:parametersContributions_IoEuropaOnly}
\end{figure*}

First, as expected, the contribution of each mutual approximation depends on the time at which it occurs. Observations collected further in time (with respect to the initial epoch $t_0$ at which the states are estimated) indeed contribute more to the initial state solution. This directly results from the fact that later observations provide tighter constraints to the initial state due to the orbit propagation: the effect of a slight variation in the initial state of Io and Europa on their trajectories grows with time. However, this time trend similarly affects both observable types and thus it has no noticeable influence on the solution improvement provided by central instants.

Nonetheless, the observable type choice also has an effect on some mutual approximations' contribution to the estimated solution. Figure \ref{fig:improvementVsContribution} displays the normalised contribution ratio of central instants over alternative observables, as defined in Section \ref{sec:contributionsDefinition}, for each mutual approximation. Some mutual approximations, mostly concentrated in the 2026-2027 period, contribute significantly less to both Io's and Europa's estimated positions when alternative observables are used instead of central instants. As expected, these observations coincide with an increase of the difference in formal errors between the two observables. The coming sections investigate why this discrepancy between the two observable types only concerns some mutual approximations and specific observational periods.

\subsection{Influence of the mutual approximations' characteristics} \label{sec:influenceMutualApproxCharac}

To better characterise the difference between the two observable types, we further analyse the relative contribution of each observation and the effect of the mutual approximation's characteristics. Section \ref{sec:influenceImpactParameterAndVelocity} discusses the influence of the impact parameter and velocity on each mutual approximation's contribution to the solution, for both central instants and alternative observables. The link between those characteristics and the observation geometry is explored in Section \ref{sec:influenceObservationGeometry}.

\subsubsection{Influence of impact parameter and velocity on each mutual approximation's contribution} \label{sec:influenceImpactParameterAndVelocity}

Focusing on the central instants case first, Figure \ref{fig:parametersContributions_centralInstants} shows that each mutual approximation's contribution to the estimated positions (averaged  between Io and Europa) strongly depends on the impact parameter and velocity. Highest contributions are systematically obtained with both low impact parameter and velocity (up to about 7 as and 1 mas/s, respectively). Mutual approximations with either low impact parameter and high impact velocity, or high impact parameter but low impact velocity also contribute significantly to the ephemerides solution. 

Using alternative observables instead of central instants alters the way some mutual approximations contribute to the estimated solution, as hinted in Section \ref{sec:directComparison20202029}. Figure \ref{fig:ratioContributionsGeometryAveraged} displays the ratio between central instants' and alternative observables' contributions to the estimated initial positions (contributions were again averaged between Io and Europa). Mutual approximations characterised by low impact parameter and high impact velocity contribute significantly less to the solution when switching to alternative observables. More precisely, most mutual approximations with impact parameters lower than 5 as and impact velocities larger than 4 mas/s contribute about 2 times more to the estimated solution when using central instants instead of alternative observables.

This analysis proves that the differences between the two observable types for some mutual approximations is amplified by specific impact characteristics. Furthermore, mutual approximations identified as unfavourable for alternative observables (low impact parameters with large impact velocities) are not evenly distributed over the 2020-2029 observational period, but rather concentrated in the 2026-2027 interval. As expected, it corresponds to the period during which the differences in formal errors between the two observables increase (Section \ref{sec:directComparison20202029}, see Figure \ref{fig:improvementVsContribution}).

It is interesting to note that mutual approximations characterised by extremely low impact parameters are also unfavourable from an observational perspective, and not only from an estimation point of view. If the two satellites eventually become so close that a (partial) occultation occurs, the observer cannot distinguish between their images anymore, introducing a gap in the apparent distance measurements near the point of closest approach and thus leading to a larger error in the estimated central instant \citep{morgado2019approx}.

\begin{figure}
	\centering
	\resizebox{\hsize}{!}{\includegraphics{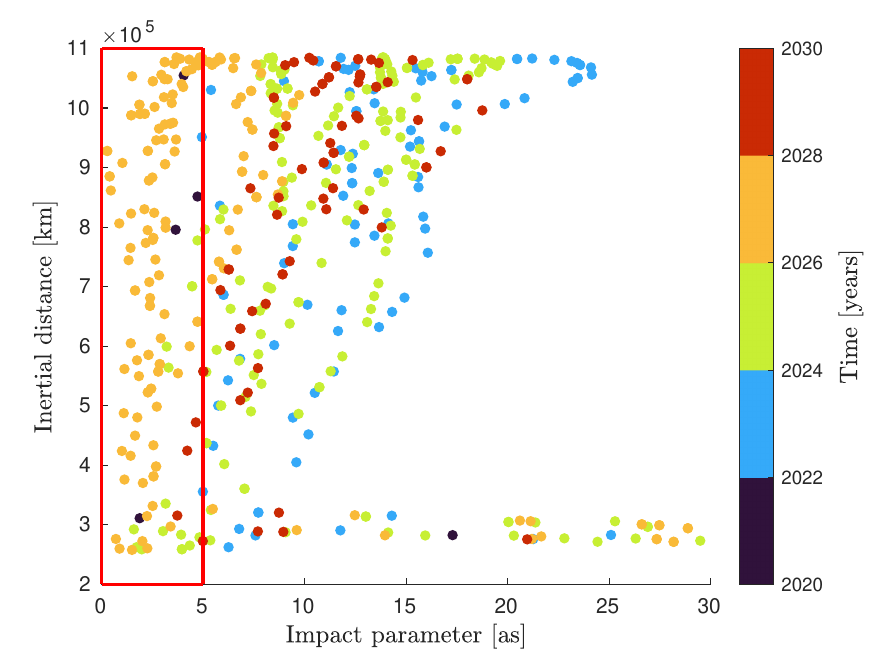}}
	\caption{Effect of the inertial geometry on the mutual approximations' impact characteristics. Both the impact parameter and the inertial distance between Io and Europa (as opposed to apparent distance as seen from the ground stations) are displayed, for each mutual approximation. The colours  represent the time at which the observation is made.}
	\label{fig:absoluteMotionIoEuropa}
\end{figure}

\begin{figure}
	\centering
	\resizebox{\hsize}{!}{\includegraphics{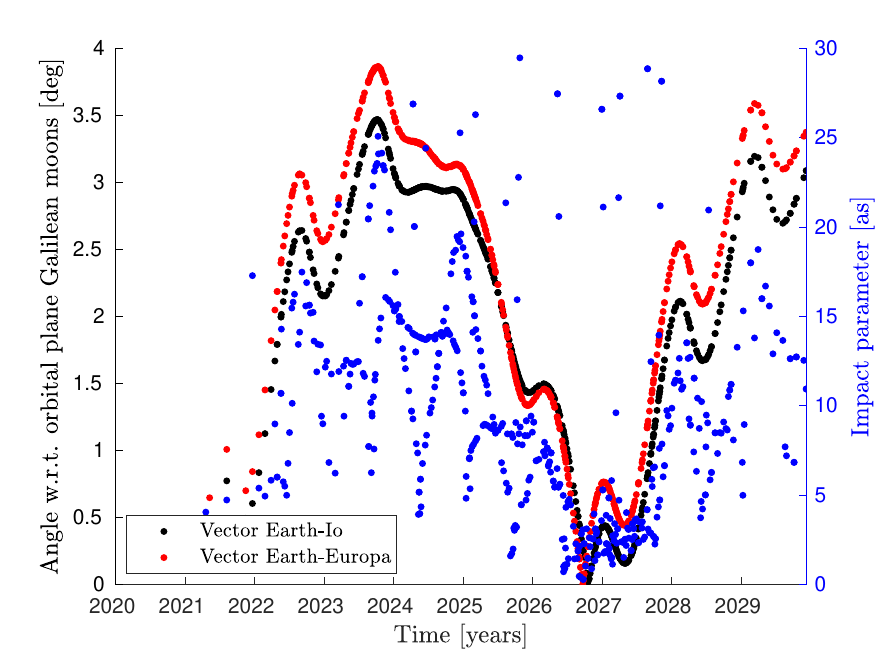}}
	\caption{Effect of the observation geometry on the mutual approximations' impact parameters. The angle between the orbital plane of the Galilean moons and the observation vectors (Earth-Io in black and Earth-Europa in red) is plotted for each mutual approximation. The corresponding impact parameters (in as) are represented by blue dots.}
	\label{fig:viewingAngleIoEuropaMutualApprox}
\end{figure}

\subsubsection{Link to the observation geometry} \label{sec:influenceObservationGeometry}

Interestingly, most of the mutual approximations simulated over the 2026-2027 period are characterised by low impact parameters (lower than 5 as), while it is not the case outside of this time interval. This is clearly shown in Figure \ref{fig:viewingAngleIoEuropaMutualApprox}, where impact parameters are displayed in blue. 

The apparent relative motion of the two moons is driven by two parameters: their inertial relative motion in the Jovian system and the observation geometry. Figure  \ref{fig:absoluteMotionIoEuropa} focuses on the former and shows the absolute distance between Io and Europa for each mutual approximation, as a function of the corresponding impact parameter. The time at which each mutual approximation occurs is displayed in colours. When excluding the 2026-2027 interval (orange dots), the impact parameters take a wide range of values (up to the limit of 30 as). The lowest impact parameters usually coincide with low inertial distances between Io and Europa, typically below $3.5\cdot10^{5}$ km (see the fraction of Figure \ref{fig:absoluteMotionIoEuropa} highlighted in red). It should be noted that the reverse is not true: low inertial distances do not automatically lead to low impact parameters.

However, during the 2026-2027 period, mutual approximations with low impact parameters are systematically achieved, even for large inertial distances between Io and Europa. This difference between the inertial and apparent relative motions is caused by the evolution of the observation geometry. Figure \ref{fig:viewingAngleIoEuropaMutualApprox} displays the angle between the orbital plane of the Galilean moons and the two observation vectors (Earth-Io and Earth-Europa), for each mutual approximation. The 2026-2027 period coincides with an almost perfect alignment between the satellite-observer vectors and the moons' orbital plane, resulting in overall lower apparent distances during Io-Europa close encounters. Those low impact parameters then regularly happen to be combined with large impact velocities. This is why mutual approximations with both aforementioned characteristics, for which the differences between the two observable types are the largest (see Section \ref{sec:influenceImpactParameterAndVelocity}), mostly occurred during years 2026 and 2027.

Such an observational period is thus not a special isolated case, but rather a periodic effect of the Earth - Jovian system geometry. Therefore, mutual approximations less favourable to alternative observables are expected to occur repeatedly, about every 6 years, and coincide with the so-called `mutual events season' during which eclipses and occultations occur. Implications of this effect of the observation geometry concerning the selection of the appropriate mutual approximations' observable type are discussed in Section \ref{sec:implications}.

It must be stressed that the selected weighting strategy for alternative observables (Section \ref{section:dataWeights}) accounts for the mutual approximation's characteristics and thus indirectly for the close encounter's geometry. The aforementioned impact of the observation geometry on the equivalence between the two observable types and more precisely on the benefit of using central instants over alternative observables is therefore already attenuated by our careful weighting of the latter. Section \ref{sec:influenceWeightingScheme} investigates the consequences of this geometry effect when not taken into consideration in the observation weights.

\subsubsection{Observational period reduced to 2026-2027} \label{sec:reducedObservationPeriod}

To quantify the impact of using a limited observation set when it unfortunately corresponds to the observational period less favourable to alternative observables, we ran additional simulations including only 2026-2027 mutual approximations in the state estimation. Table \ref{tab:formalErrors20262027Observations} compares the improvement in formal errors provided by central instants over alternative observables with either complete (2020-2029) or partial (2026-2027) observation sets. 

As expected, the differences in formal errors between the two observables increase when only considering 2026-2027 observations, except for Europa's axial position component. The improvement provided by central instants is multiplied by factors ranging from 1.5 to 3 for most position and velocity components. Compared to results obtained with the complete observation set, formal errors' improvements in the axial direction become more significant. They are even comparable to those achieved in the radial direction for both Io's and Europa's velocity. 

\begin{table}[h]
	\caption{Improvements in the final errors obtained with central instants, using two different observations sets.}
	\label{tab:formalErrors20262027Observations}
	\centering
	\begin{tabular}{l  l  c c c}
		\hline \hline \noalign{\vskip 0.02in}
		\multicolumn{2}{l}{Parameters } & \multicolumn{2}{c}{Improvement formal errors} & Ratio \\
		\multicolumn{2}{c}{ } & 2020-2029 [1] & 2026-2027 [2] & [2]/[1] \\ \hline \noalign{\vskip 0.02in}
		Position &radial & 21.9 \% & 45.2 \% & 2.1  \\
		Io &normal & 17.4 \% & 26.7 \% &  1.5 \\
		&axial & 12.0 \% & 29.5 \% & 2.5 \\ \hline \noalign{\vskip 0.02in}
		Velocity& radial &17.3 \% & 26.5 \%& 1.5\\
		Io &normal & 22.3 \% & 46.3 \% & 2.1\\
		&axial & 2.9 \% & 25.1 \% & 8.6\\ \hline \noalign{\vskip 0.02in}
		Position&radial &19.7 \% & 59.8 \% & 3.0 \\
		Europa &normal&20.4 \% & 27.8 \% & 1.4 \\
		&axial & 11.0 \% & 2.5 \% & 0.2 \\ \hline \noalign{\vskip 0.02in}
		Velocity&radial &22.1 \% & 29.3 \% & 1.3\\
		Europa &normal & 19.7 \% & 59.7 \% & 3.0\\
		&axial & 4.2 \% & 33.6 \% & 7.8\\ \hline                     
	\end{tabular}
	\tablefoot{The improvement in formal errors obtained with central instants is computed with respect to formal errors resulting from alternative observables, in two different cases. First, all mutual approximations simulated over the whole 2020-2029 period are included (referred to as case [1] in the table). Second, only mutual approximations occurring during the 2026-2027 period are selected (referred to as [2]).}
\end{table}

Although alternative observables and central instants were proven to lead to very comparable solutions in nominal configurations, the influence of the available set of observations must thus not be neglected. If the number of mutual approximations is limited, or the period over which they were observed too short, it is recommended to investigate the characteristics of the available mutual approximations before selecting an observable type. The improvement provided by using central instants is indeed amplified when exclusively including mutual approximations observed under unfavourable observation angles (2026-2027 period in our case).

\subsection{Implications} \label{sec:implications}

If enough observations are available, the improvement provided by using central instants instead of alternative observables is limited to about 20\% for our Io-Europa test case. However, such improvement might still be relevant when concurrently estimating other dynamical parameters along with natural bodies' ephemerides. Accurate determination of the tidal dissipation, in particular, is required to gain insights into the orbital evolution of planetary systems. 

Recent estimations from astrometric data indeed showed that Saturn's tidal quality factor $Q$ can vary by several orders of magnitude from one moon's forcing frequency to another \citep{lainey2020resonance}. These results are highly inconsistent with current evolution models and thus they highlight the need for accurate frequency-dependent estimation of tidal parameters \citep{fuller2016resonance}. This is for instance not yet done for Jupiter, whose tidal quality factor is currently only determined at Io's frequency. A 20\% improvement in the formal errors of the natural satellites' initial states might be critical for such applications. As many perturbing dynamical effects can be absorbed by a variation in the initial state, any improvement in the state estimation can indeed help to detect and estimate such small tidal effects.

Furthermore, a 20\% improvement in the predicted position and/or velocity of the targeted body is not negligible for orbital design applications. It can indeed affect the corrective manoeuvres required before and after a flyby or an orbital insertion, allowing for a more efficient design and thus reducing the $\Delta$V budget.

\begin{figure*} 
	\centering
	\begin{subfigure}{0.49\textwidth}
		\centering
		{\includegraphics[width=1.0\textwidth]{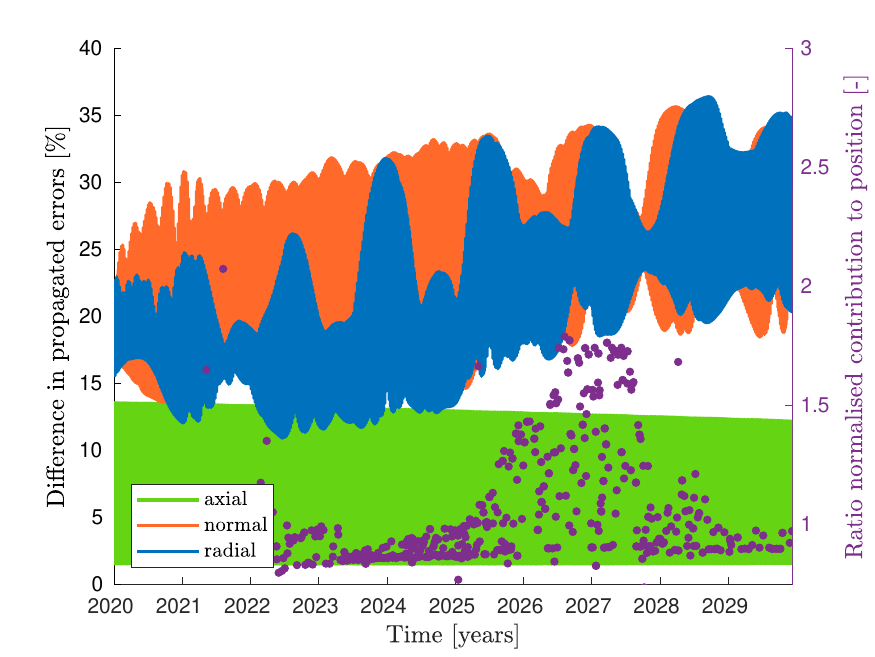}}
		\caption{}
		\label{fig:propagatedErrorsIo}
	\end{subfigure}
	\begin{subfigure}{0.49\textwidth}
		\centering
		{\includegraphics[width=1.0\textwidth]{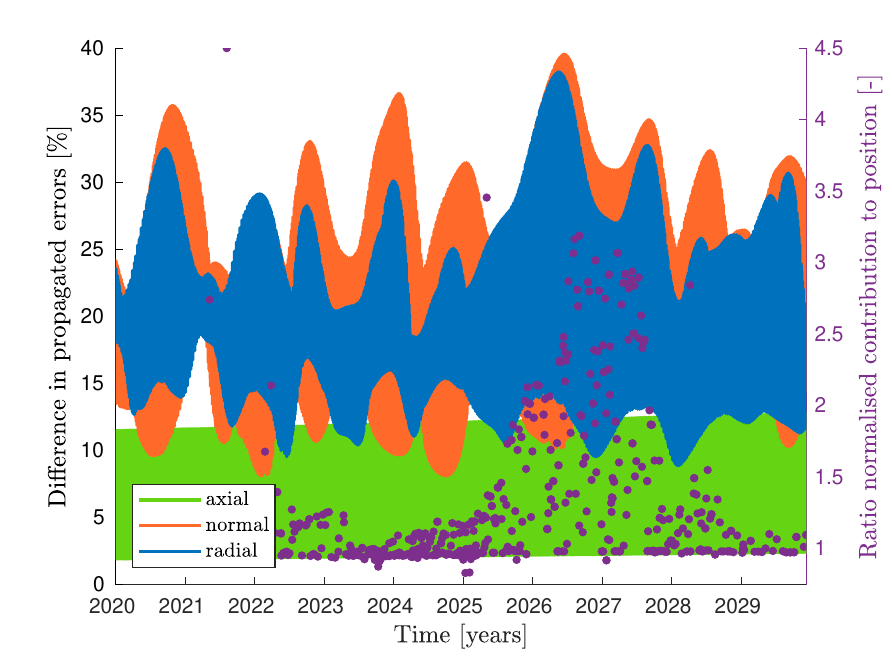}}
		\caption{}
		\label{fig:propagatedErrorsEuropa}
	\end{subfigure}
	\caption{Propagated formal errors in position components over the observational period (panel a: Io, panel b: Europa). The errors in position are expressed in the RSW frame (see Figure \ref{fig:referenceFrames}) and are obtained from the propagated covariance matrix and estimated initial states. The purple dots (right axis) display the ratio between the normalised contributions of central instants over their corresponding alternative observables, for each mutual approximation.}
	\label{fig:propagatedErrors}
\end{figure*}

The impact of the mutual approximations' observables choice then depends on the timing of the manoeuvre. Figure \ref{fig:propagatedErrors} displays the improvement in propagated errors in radial, normal, and axial direction for both Io's and Europa's positions. The 20\% difference in the formal errors of the initial states can increase once propagated, at least in the radial and normal directions. Depending on the time at which the manoeuvre is planned, the improvement in the accuracy of the predicted targeted body's state might thus be higher than 20\% if central instants are used. Looking at Figure \ref{fig:propagatedErrorsIo}, differences can reach up to 35\% for Io's radial and normal position components while they increase up to almost 40\% for Europa (see Figure \ref{fig:propagatedErrorsEuropa}). 

For Europa, the largest differences in propagated errors clearly correspond to the 2026-2027 period and thus coincide with the least favourable observation geometry for alternative observables (see discussion in Section \ref{sec:influenceMutualApproxCharac}). This effect is however barely noticeable in Figure \ref{fig:propagatedErrorsIo}, for Io's case. Yet, it should be pointed that when a very sensitive manoeuvre is planned during such a particular observational period, the impact of the observable choice on the quality of the estimated solution might not be negligible.

All aforementioned points are direct consequences of the imperfect equivalence between the two types of observables, which might be accentuated when fewer observations are available. In practice, however, this would be balanced by other observations combined with mutual approximations in the state estimation. Building on  Section \ref{sec:influenceMutualApproxCharac}, it is nonetheless worth highlighting that the number and distribution of the observations should be carefully considered when selecting an appropriate observable for the processing of mutual approximations data.

The fact that the alternative observables' unfavourable observational period corresponds to the mutual events season might also be an interesting finding for the processing of the mutual events themselves. Indeed, if the timings at which eclipses and occultations occur were to be directly used as observables (as it is the case with mutual approximations' central instants), the aforementioned effect of the observation geometry would first need to be carefully investigated.

\subsection{Influence of the weighting scheme} \label{sec:influenceWeightingScheme}

When enough mutual approximations are available, the good match which can be achieved between both observable types' estimated solutions actually strongly depends on the accurate weighting of each mutual approximation. As highlighted in Section \ref{section:dataWeights}, individual weights have to be computed for alternative observables. Using a single averaged value for alternative observables' weights yields a much larger discrepancy with respect to the central instants solution over the full simulation, as reported in Table \ref{tab:influenceWeighting}. Compared to formal errors obtained with alternative observables, those achieved with central instants are then reduced by a factor 1.5 to 2.7 for the initial position components, and up to a factor 4 for the velocity components. 

These results clearly prove that the equivalence  between the two observable types is conditioned by the appropriate weighting of alternative observables. As mentioned in Section \ref{section:dataWeights}, it is crucial to carefully compute suitable observation weights to ensure that the alternative observables indirectly take into account the geometry of the close encounter in the plane of the sky. 

\cite{emelyanov2017precision} already conducted a comparative analysis between the two types of observables, although the central instant partials were computed numerically. Interestingly, a convergence issue was encountered with alternative observables, while none was reported for central instants. This indicates that the two observables types were not perfectly equivalent in this case, which might be due to the weighting issue we just highlighted.

\begin{table}[h]
	\caption{Comparison of the final formal errors obtained with the two types of observables, when applying constant weight values.}
	\label{tab:influenceWeighting}
	\centering
	\begin{tabular}{l c c c}
		\hline \hline \noalign{\vskip 0.02in}
		Parameters & \multicolumn{2}{c}{Formal errors with} & Ratio \\
		& Central & Alternative &  \\
		& instants [1] & observables [2] & [2]/[1] \\ \hline \noalign{\vskip 0.02in}
		$x^{\prime}_{\mathrm{Io}}$ & 7.01 km & 11.2 km & 1.6 \\
		$y^{\prime}_{\mathrm{Io}}$ & 10.4 km & 17.6 km & 1.7 \\
		$z^{\prime}_{\mathrm{Io}}$ & 14.5 km & 37.3 km & 2.6 \\ \noalign{\vskip 0.02in} \hline \noalign{\vskip 0.02in}
		$\dot{x}^{\prime}_{\mathrm{Io}}$ & 0.372 m/s & 1.30 m/s & 3.5 \\
		$\dot{y}^{\prime}_{\mathrm{Io}}$ & 0.373 m/s & 0.904 m/s & 2.4 \\
		$\dot{z}^{\prime}_{\mathrm{Io}}$ & 0.562 m/s & 1.86 m/s & 3.3 \\ \noalign{\vskip 0.02in} \hline \noalign{\vskip 0.02in}
		$x^{\prime}_{\mathrm{Europa}}$ & 2.31 km & 6.24 km & 2.7 \\
		$y^{\prime}_{\mathrm{Europa}}$ & 13.6 km & 20.5 km & 1.5 \\
		$z^{\prime}_{\mathrm{Europa}}$ & 26.3 km & 49.1 km & 1.9 \\ \noalign{\vskip 0.02in} \hline \noalign{\vskip 0.02in}
		$\dot{x}^{\prime}_{\mathrm{Europa}}$ & 0.191 m/s & 0.535 m/s & 2.8 \\
		$\dot{y}^{\prime}_{\mathrm{Europa}}$ & 0.222 m/s & 0.851 m/s & 3.8 \\
		$\dot{z}^{\prime}_{\mathrm{Europa}}$ & 0.454 m/s & 1.75 m/s & 3.9 \\ \hline                            
	\end{tabular}
	\tablefoot{The final errors are expressed in satellites' central body-centred cartesian coordinates. The central instants solution is obtained with a constant weight $\sigma_{t_c}$ = 3.5 s (see Section \ref{section:dataWeights}). For the alternative observables, we use the average of the individual weights defined by Equation \ref{eqn:alternativeObservableWeight}: $\sigma_{\mathrm{alt.}} = 8.87\times 10^{-3}$ mas/s.}
\end{table}

\subsection{Verification case: Four Galilean moons} \label{sec:validationFourGalileanMoons}

As verification, we simulated mutual approximations between the four Galilean moons of Jupiter. Ganymede's and Callisto's initial states were then estimated in addition to Io's and Europa's. The resulting formal errors in the four moons' initial states are provided in Table \ref{tab:formalErrorsFourMoons}. Even if not displayed here, the time evolution of the errors in Io's and Europa's initial states show a behaviour comparable to what was observed in the first test case limited to Io and Europa only (see Figure \ref{fig:formalErrorsHistoryIoEuropaOnly}). The final formal errors are however a bit lower in the four moons' case. 

\begin{table}[h]
	\caption{Comparison of the formal errors in the initial positions of the four Galilean moons.}
	\label{tab:formalErrorsFourMoons}
	\centering
	\begin{tabular}{l l  c c  c}
		\hline \hline \noalign{\vskip 0.02in}
		\multicolumn{1}{l}{Moon} & Position & \multicolumn{2}{c}{Formal errors [km]} & Relative \\
		& component &Central & Alternative & difference  \\
		\multicolumn{2}{l}{ } & instants & observables & [\%] \\ \hline \noalign{\vskip 0.02in}
		& Radial & 0.222 & 0.242 & 8.2 \\
		Io& Normal & 6.60 & 7.30 & 9.6 \\
		& Axial & 4.49 & 4.57 & 1.7 \\ \hline \noalign{\vskip 0.02in}
		& Radial & 0.908 &  1.02 & 10.7 \\
		Europa & Normal & 5.27 & 5.90 & 10.7 \\
		& Axial & 10.3 & 10.1 & 2.2 \\ \hline \noalign{\vskip 0.02in}
		& Radial & 1.22 & 1.37 & 11.3 \\
		Ganymede & Normal & 5.99 & 6.39 & 6.2 \\
		& Axial & 17.9 & 18.1 & 1.1 \\ \hline \noalign{\vskip 0.02in}
		& Radial & 1.55 & 1.56 & 0.7\\
		Callisto & Normal & 8.64 & 8.93 & 3.2 \\
		& Axial & 25.5 & 26.0 & 1.6 \\ \hline                            
	\end{tabular}
	\tablefoot{The formal errors in position are expressed in the RSW frame (see Figure \ref{fig:referenceFrames}) and are provided for the two types of observables. The last column displays the relative difference between the two. All predicted mutual approximations between the four moons over the period 2020-2029 were included in the state estimation.}
\end{table}

Similarly to the Io-Europa test case, using central instants over alternative observables mostly improves errors in the radial and normal directions, but only has a marginal effect on the axial direction errors. For Europa's axial position, the estimation is even 2\% better with alternative observables. As shown in Table \ref{tab:formalErrorsFourMoons}, differences in formal errors between the two types of observables are significantly lower for Callisto than for the three other moons. These differences are only of a few percents, while they reach about 10\% for Io's, Europa's, and Ganymede's initial positions. 

\begin{figure*}
	\centering
	\includegraphics[width=17cm]{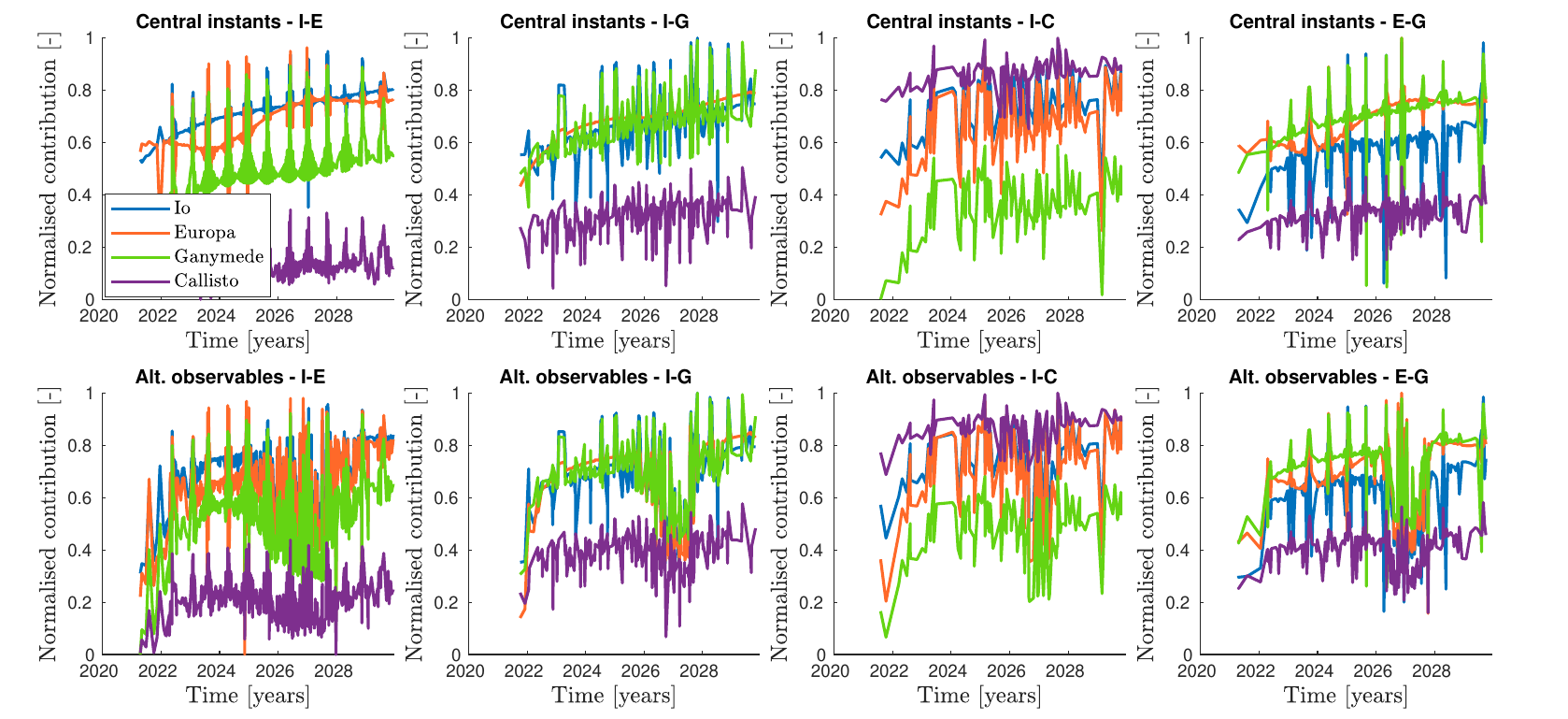}
	\caption{Normalised contribution of mutual approximations between Io and Europa (I-E, left plots), Io and Ganymede (I-G, centre-left plots), Io and Callisto (I-C, centre-right plots), Europa and Ganymede (E-G, right plots) to the initial position of Io (blue), Europa (orange), Ganymede (green), and Callisto (purple). Results for Europa-Callisto and Ganymede-Callisto mutual approximations are not represented here, but do not show any trend that is not already highlighted by the contributions of the other observations. The first simulated observation only happens at the end of year 2021, explaining the lack of data before that date.}
	\label{fig:contributionsFourGalileanMoons}
\end{figure*}

Again, the contribution of each mutual approximation to the estimated ephemerides solution also reflects the higher sensitivity of central instants to the complex dynamics at play in the Jovian system. Figure \ref{fig:contributionsFourGalileanMoons} shows the contribution of every observation to the initial position of Io (blue), Europa (orange), Ganymede (green), and Callisto (purple). As expected, all mutual approximations contribute to estimating the initial states of Io, Europa, and Ganymede because of the Laplace resonance between these three moons. On the other hand, only mutual approximations directly involving Callisto significantly help to determine its initial state.

In Figure \ref{fig:contributionsFourGalileanMoons}, clear periodic patterns can be identified in the central instants case, at least when enough observations are available, such as for Io-Europa, Io-Ganymede, and Europa-Ganymede mutual approximations. While still present, those patterns are however less pronounced for the alternative observables case. They are directly related to the relative motion of the satellites in the Jovian system (inertial motion, as opposed to apparent). This again indicates that part of the information encoded in mutual approximations is not fully captured by alternative observables. 

As already highlighted by the Io-Europa case (Section \ref{sec:directComparison20202029}), central instants are indeed more sensitive to the apparent relative acceleration between the two satellites. Central instants' partials directly account for any acceleration variation induced by a small change in initial states, while alternative observables do not. This may also explain why differences between the two observables are lower for Callisto. As it is the most remote moon with respect to Jupiter, the distance between Callisto and each of the three other moons is larger than the distances between Io, Europa, and Ganymede, resulting in smaller inter-moon accelerations. Furthermore, because of the Laplace resonance between the three innermost Galilean satellites, accelerations exerted by one of these three moons on the other two strongly influence their dynamics. These two combined effects might strengthen the advantage of central instants over alternative observables for Io, Europa, and Ganymede, compared to Callisto.

Overall, all findings obtained in the first simple Io-Europa test case are confirmed by this second analysis extended to the four Galilean moons. The two types of observables lead to almost equivalent solutions if the alternative observables are properly weighted (see Section \ref{section:dataWeights}), despite a 10\% reduction of the formal errors in both radial and normal directions when using central instants. The effect of the observation geometry is also similar to the Io-Europa case: low impact parameter, large impact velocity mutual approximations simulated in 2026-2027 are unfavourable to alternative observables.

\section{Conclusion \& discussion} \label{sec:conclusions}
We developed an analytical formulation for the observation partials of the mutual approximations' central instants. This allows those central instants to be directly used as observables to estimate the ephemerides of natural satellites. Our analytical method relies on  a second-order polynomial to approximate the relative motion of two satellites around their point of closest approach. From this polynomial function, we derived an expression for the central instant as a function of the apparent relative position, velocity, and acceleration of the two satellites. 

Higher-order terms could theoretically be included in our formulation. Using a third-order polynomial to reproduce the apparent relative motion of the two moons (Equations \ref{eqn:polynomialFunctionX}-\ref{eqn:polynomialFunctionY}) would lead to a fourth-order polynomial for the central instant (Equation \ref{eqn:cubicPolynomial}). The roots of such a polynomial could still be computed analytically, but at the cost of a dramatic increase in complexity. However, a second-order polynomial has been proven sufficient to capture the apparent relative dynamics of the two satellites around their closest encounter and to yield highly accurate analytical partials for central instants. 

Numerically computing partials for central instants is extremely computationally demanding \citep{emelyanov2017precision}. It requires to independently propagate small variations in each of the estimated parameters (at least 6 initial state components for each of the two moons involved). Afterwards, the new central instants must be determined, which is a time-consuming process in itself. Should central instants be used, our analytical approach is thus significantly faster than the numerical computation of the observation partials. 

We conducted a comparative covariance analysis using either only central instants, or only mutual approximation's alternative observables to estimate the Galilean moons' ephemerides. When using the entire set of viable mutual approximations over the period 2020-2029, the difference in  formal errors does not exceed 20\%  between the two types of observables. The central instants achieve the best ephemerides solution because alternative observables do not account for some of the dynamical effects affecting the close encounter (e.g. apparent relative acceleration between the two satellites). In contrast, these effects are directly captured by central instants, which is beneficial for the resulting estimated solution.

Overall, we still prove alternative observables to be almost equivalent to central instants, but only under specific conditions. First, when using alternative observables, the shape of the observed close encounter must indirectly be accounted for in the calculation of the observation weights, while it is automatically included in the central instant case. Individual and accurate weighting of each event, based on the apparent relative dynamics of the satellites, is then crucial to obtain a consistent solution. It is indeed necessary to convert any error in the estimated central instant to an error in the derivative of the apparent distance. We show that when using a single averaged value to weight all mutual approximations, the formal errors in initial states obtained with central instants were 1.5 to 4 times lower than with alternative observables in our test case. As discussed in Section \ref{sec:influenceWeightingScheme}, an inappropriate weighting scheme could thus possibly explain previous indications of a non-equivalence between the two observable types \citep{emelyanov2017precision}. When using alternative observables, we therefore recommend to adopt the weighting strategy described in Section \ref{section:dataWeights}, and more precisely to compute the weights with Equation \ref{eqn:alternativeObservableWeight}. In Appendix \ref{appendix:weightsPastObservations}, we provide the appropriate alternative observables' weights for the 2016-2018 mutual approximations reported in \cite{morgado2019approx}. These weight values should be applied for the 2016-2018 observations to be properly included in the state estimation. 

Furthermore, all mutual approximations do not homogeneously contribute to the ephemerides solution. The satellites' dynamics are overall better constrained by mutual approximations with a low impact parameter (typically below 7 as) and low relative velocity (1 mas/s), for both observable types. However, some characteristics in particular are unfavourable to alternative observables: mutual approximations with low impact parameters but large impact velocities contribute significantly more to the estimated solution when using central instants (factor 2 to 3). Preferring central instants is thus particularly advantageous for these specific mutual approximations, which are not isolated events but periodically represent most of the observations for one or two years (see discussion in Section \ref{sec:influenceObservationGeometry}).

Choosing between the two types of observables when estimating ephemerides from mutual approximations therefore requires critical evaluation. If many mutual approximations are available to estimate the moons' ephemerides, one can safely use alternative observables without substantially degrading the solution. However, this does not systematically hold for a small set of observations, especially if they are all collected during the alternative observables' unfavourable observation period. The formal error reduction provided by our method then strongly depends on the mutual approximations' characteristics. 

The relevance of selecting central instants over alternative observables eventually depends on the application of the ephemerides solution. As detailed in Section \ref{sec:implications}, a 10-20\% improvement in the formal errors of the satellites' state might be significant when concurrently estimating tidal parameters. It may also be non-negligible for mission design applications. Improved ephemerides are indeed crucial to design efficient flybys or orbital insertions requiring only limited corrective manoeuvres. The timing of the manoeuvres must then be taken into consideration to select a suitable observable, especially if they coincide with observation geometries less favourable to alternative observables.

As mentioned in Section \ref{sec:implications}, a comparable analysis could be conducted for mutual events. We expect to obtain consistent results with respect to the mutual approximations' case, given the similarities between the two types of observation. However, all mutual events occur during the alternative observables' unfavourable observational period. It is thus important to confirm the influence of the observation geometry on the differences between central instants' and alternative observables' state estimation solutions. This could be an interesting result, in case the timings of eclipses and occultations would be directly used as observables, as for mutual approximations.

\begin{acknowledgements} 
	We would like to thank Wouter van der Wal for his useful feedback, which significantly improved the quality of this manuscript.
\end{acknowledgements}

\bibliographystyle{aa} 
\bibliography{references}

\appendix

\section{Fitting a polynomial to a close encounter's apparent distance history} \label{appendix:polynomialFitting}

\setcounter{figure}{0}
\setcounter{table}{0}

\begin{figure*}
	\centering
	\includegraphics[width=17cm]{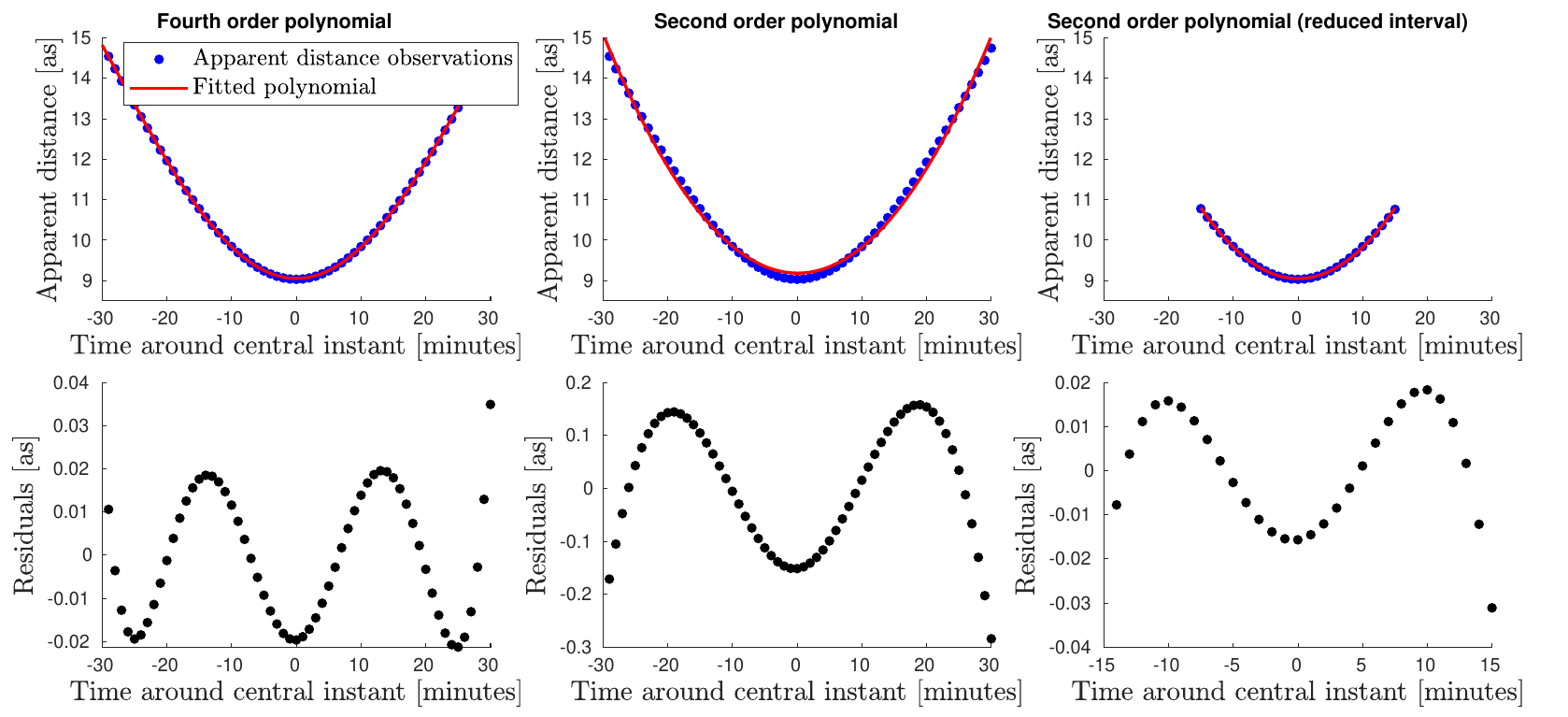}
	\caption{Apparent distance measurements during a mutual approximation (blue dots on top panels). The polynomial used to fit these observations is typically a fourth order one (left panel). A second order polynomial was also tested, for the whole duration of the event (middle panels) and over a reduced time interval (30 minutes) centred on the central instant $t_c$. For each of the three case, the residuals between the fitted polynomial and the true apparent distance history are displayed in the bottom panels.}
	\label{fig:influenceOrderFittingPolynomial}
\end{figure*}

As described in Section \ref{sec:introduction}, the apparent distance history during a close encounter between the two satellites is typically fitted with a fourth order polynomial \citep[e.g.][]{morgado2016astrometry}. This allows to estimate the mutual approximation's central instant, as well as its impact parameter for instance. In this appendix, we discuss the influence of the order of the fitting polynomial. The maximum absolute values of the residuals between the fitted polynomial and the true apparent distance history are reported in Table \ref{tab:influenceOrderFittingPolynomial} (for the first mutual approximations predicted between Io and Europa, starting from 01/01/2020). For clarity, the apparent distance observations, fitted polynomial and resulting residuals are displayed in Figure \ref{fig:influenceOrderFittingPolynomial} for the first mutual approximation.

When switching from a fourth order to a second order polynomial to reproduce the apparent distance history over the whole duration of the close encounter (i.e. 60 minutes), the residuals increase by almost a factor 10. However, if we only focus on a small fraction of the event (here only 30 minutes centred on the central instant), a second order polynomial achieves similar residuals as a fourth order polynomial applied to the full close encounter duration. This proves that a second order polynomial is well-suited when focusing on short time intervals centred on $t_c$. This is the case when deriving observation partials for central instants, as we then only consider slight changes in $t_c$, induced by small variations of the estimated parameters.

\begin{table}[h]
	\caption{Maximum absolute values for the residuals between the apparent distance history and the fitted polynomial.}
	\label{tab:influenceOrderFittingPolynomial}
	\centering
	\begin{tabular}{c c  c  c}
		\hline \hline \noalign{\vskip 0.02in}	
		Mutual & \multicolumn{3}{c}{Max. absolute residual value [as]} \\
		approx. & \multicolumn{2}{c}{$[t_c-30 \mathrm{min};t_c + 30 \mathrm{min}]$} & $[t_c-15 \mathrm{min};t_c + 15 \mathrm{min}]$ \\
		& $4^{\mathrm{th}}$ order & $2^{\mathrm{nd}}$ order &  $2^{\mathrm{nd}}$ order \\ \hline \noalign{\vskip 0.02in}	
		1 & $3.50\cdot 10^{-2}$ & $2.84\cdot 10^{-1}$ & $3.11\cdot 10^{-2}$\\
		2 & $3.13\cdot 10^{-2}$ & $2.45\cdot 10^{-1}$ & $2.97\cdot 10^{-2}$\\
		3 & $3.55\cdot 10^{-2}$ & $2.67\cdot 10^{-1}$ & $3.28\cdot 10^{-2}$\\
		4 & $3.84\cdot 10^{-2}$ & $2.99\cdot 10^{-1}$  & $3.40\cdot 10^{-2}$\\
		5 & $3.78\cdot 10^{-2}$ & $2.79\cdot 10^{-1}$ & $3.32\cdot 10^{-2}$\\
		\hline                         
	\end{tabular}
	\tablefoot{The residuals are compared between three configurations: fourth order and second order polynomials used over the whole duration of the close encounter (i.e. 60 minutes: $[t_c-30 \mathrm{min};t_c + 30 \mathrm{min}]$) and second order polynomial over a reduced interval centred on the central instant (only 30 minutes: $[t_c-15 \mathrm{min};t_c + 15 \mathrm{min}]$). Results are reported for 5 mutual approximations.}
\end{table}

\section{Position and velocity partials of $\alpha_{_{Si}}$, $\delta_{_{Si}}$, $\dot{\alpha}_{_{Si}}$, $\dot{\delta}_{_{Si}}$, $\ddot{\alpha}_{_{Si}}$ and $\ddot{\delta}_{_{Si}}$} \label{appendix:partials}

\subsection{$\alpha_{_{Si}}$ and $\delta_{_{Si}}$ partials}

First, we derive the partials of the right ascension $\alpha_{_{Si}}$ and declination $\delta_{_{Si}}$ with respect to the two satellites' and observer's positions, as follows:
\begin{align}
\left.\frac{\partial \alpha_{_{Si}}}{\partial \boldsymbol{r}_{_{Si}}}\right|&_{t_{_{Si}}} = - \left.\frac{\partial \alpha_{_{Si}}}{\partial \boldsymbol{r}_{_{O}}}\right|_{t_{_{Si}}} =  \frac{1}{r_{i_{xy}}^2} \begin{pmatrix}
-y_i \\ x_i \\ 0
\end{pmatrix},  \\
\left.\frac{\partial \delta_{_{Si}}}{\partial \boldsymbol{r}_{_{Si}}}\right|&_{t_{_{Si}}} = - \left.\frac{\partial \delta_{_{Si}}}{\partial \boldsymbol{r}_{_{O}}}\right|_{t_{_{Si}}} = \frac{1}{r_{i}^2 r_{i_{xy}}} \begin{pmatrix}
-x_i z_i \\ - y_i z_i \\ r_{i_{xy}}^2 \end{pmatrix},  
\end{align}
\begin{align}
\left.\frac{\partial \left[\alpha,\delta\right]_{_{Si}}}{\partial \boldsymbol{r}_{_{Sj}}}\right|&_{t_{_{Si}}} = \begin{pmatrix}
0 \\ 0 \\ 0 \end{pmatrix} ; i \neq j.
\end{align}
The partials of $\alpha_{_{Si}}$ and $\delta_{_{Si}}$ with respect to the velocity vectors are by definition equal to zero:
\begin{align}
\left.\frac{\partial \left[\alpha,\delta\right]_{_{Si}}}{\partial \dot{\boldsymbol{r}}_{_{Si}}}\right|_{t_{_{Si}}} & = \left.\frac{\partial \left[ \alpha,\delta\right]_{_{Si}}}{\partial \dot{\boldsymbol{r}}_{_{O}}}\right|_{t_{_{Si}}} = \left.\frac{\partial \left[\alpha,\delta\right]_{_{Si}}}{\partial \dot{\boldsymbol{r}}_{_{Sj}}}\right|_{t_{_{Si}}} = \begin{pmatrix}
0 \\ 0 \\ 0 \end{pmatrix} ; i \neq j. 
\end{align}

\subsection{$\dot{\alpha}_{_{Si}}$ and $\dot{\delta}_{_{Si}}$ partials}

The partials of $\dot{\alpha}_{_{Si}}$ and $\dot{\delta}_{_{Si}}$ with respect to the position vectors of the two satellites and the observer are
\begin{align}
\left.\frac{\partial \dot{\alpha}_{_{Si}}}{\partial \boldsymbol{r}_{_{Si}}}\right|_{t_{_{Si}}} &= \frac{1}{r_{i_{xy}}^4}\begin{pmatrix}
\dot{y}_i\left(y_i^2 - x_i^2\right)+2x_i y_i \dot{x}_i \\ \addlinespace
\dot{x}_i\left(y_i^2 - x_i^2\right)-2x_i y_i \dot{y}_i \\ \addlinespace 0
\end{pmatrix}, \\
\left.\frac{\partial \dot{\delta}_{_{Si}}}{\partial \boldsymbol{r}_{_{Si}}}\right|_{t_{_{Si}}} &= \frac{1}{r_{i}^2r_{i_{xy}}}\begin{pmatrix}
-z_i\dot{x}_i +x_i\dot{z}_i\\ \addlinespace -z_i\dot{y}_i +y_i\dot{z}_i\\ \addlinespace
-x_i \dot{x}_i -y_i \dot{y}_i
\end{pmatrix} -\frac{2\dot{z}_ir_{i_{xy}}}{r_{i}^4}\begin{pmatrix}x_i\\ \addlinespace y_i\\ \addlinespace z_i\end{pmatrix} \nonumber \\
& + \frac{z_i\left(x_i \dot{x}_i +y_i \dot{y_i}\right)}{r_{i}^4r_{i_{xy}}}\left[\left(\frac{z_i^2}{r_{i_{xy}}^2}+3\right)\begin{pmatrix}
x_i \\ \addlinespace y_i \\ \addlinespace 0
\end{pmatrix} + \begin{pmatrix}
0 \\ \addlinespace 0 \\ \addlinespace z_i
\end{pmatrix}\right], \\
\left.\frac{\partial [\dot{\alpha},\dot{\delta}]_{_{Si}}}{\partial \boldsymbol{r}_{_{O}}}\right|_{t_{_{Si}}} &= - \left.\frac{\partial [\dot{\alpha},\dot{\delta}]_{_{Si}}}{\partial \boldsymbol{r}_{_{Si}}}\right|_{t_{_{Si}}}, \\
\left.\frac{\partial [\dot{\alpha},\dot{\delta}]_{_{Si}}}{\partial \boldsymbol{r}_{_{Sj}}}\right|_{t_{_{Si}}} &=  \begin{pmatrix}
0 \\ 0 \\ 0 \end{pmatrix} ; i \neq j.
\end{align}
We also compute partials of $\dot{\alpha}_{_{Si}}$ and $\dot{\delta}_{_{Si}}$ with respect to the two satellites' and the observer's velocity vectors:
\begin{align}
\left.\frac{\partial \dot{\alpha}_{_{Si}}}{\partial \dot{\boldsymbol{r}}_{_{Si}}}\right|_{t_{_{Si}}} &= - \left.\frac{\partial \dot{\alpha}_{_{Si}}}{\partial \dot{\boldsymbol{r}}_{_{O}}}\right|_{t_{_{Si}}} = \frac{1}{r_{i_{xy}}^2} \begin{pmatrix}
-y_i \\ x_i \\ 0
\end{pmatrix},  \\
\left.\frac{\partial \dot{\delta}_{_{Si}}}{\partial \dot{\boldsymbol{r}}_{_{Si}}}\right|_{t_{_{Si}}} &= - \left.\frac{\partial \dot{\delta}_{_{Si}}}{\partial \dot{\boldsymbol{r}}_{_{O}}}\right|_{t_{_{Si}}} = \frac{1}{r_{i}^2 r_{i_{xy}}} \begin{pmatrix}
-x_i z_i \\ - y_i z_i \\ r_{i_{xy}}^2 \end{pmatrix}, \\
\left.\frac{\partial \left[\dot{\alpha},\dot{\delta}\right]_{_{Si}}}{\partial \dot{\boldsymbol{r}}_{_{Sj}}}\right|_{t_{_{Si}}} &=  \begin{pmatrix}
0 \\ 0 \\ 0 \end{pmatrix} ; i \neq j.
\end{align}

\subsection{$\ddot{\alpha}_{_{Si}}$ and $\ddot{\delta}_{_{Si}}$ partials}

The partials of $\ddot{\alpha}_{_{Si}}$ and $\ddot{\delta}_{_{Si}}$ with respect to position vectors lead to more complex expressions. We therefore split those partials into two terms. The first one, denoted as $g_{\ddot{\alpha}_{_{Si}}}$ or $g_{\ddot{\delta}_{_{Si}}}$, correspond to the contribution of the acceleration partials (more details about how to compute them are provided in Appendix \ref{appendix:accelerationPartials}). The rest of the partial expression is included in the other term ($g^{\prime}_{\ddot{\alpha}_{_{Si}}}$ or $g^{\prime}_{\ddot{\delta}_{_{Si}}}$). 

We thus obtain the following formulation for the partial of $\ddot{\alpha}_{_{Si}}$ with respect to the position of satellite $i$:
\begin{align}
\left.\frac{\partial \ddot{\alpha}_{_{Si}}}{\partial  \boldsymbol{r}_{_{Si}}}\right|_{t_{_{Si}}} &= g_{\ddot{\alpha}_{_{Si}}} + g^{\prime}_{\ddot{\alpha}_{_{Si}}}, \text{ with } \\ 
g_{\ddot{\alpha}_{_{Si}}} &= \begin{pmatrix}
x_i \frac{\partial \ddot{y}_i}{\partial x_{_{Si}}} - y_i \frac{\partial \ddot{x}_i}{\partial x_{_{Si}}} \\ \addlinespace
x_i \frac{\partial \ddot{y}_i}{\partial y_{_{Si}}} - y_i \frac{\partial \ddot{x}_i}{\partial y_{_{Si}}}
\\ \addlinespace x_i \frac{\partial \ddot{y}_i}{\partial z_{_{Si}}} - y_i \frac{\partial \ddot{x}_i}{\partial z_{_{Si}}}
\end{pmatrix}, \\
g^{\prime}_{\ddot{\alpha}_{_{Si}}}
&= \begin{pmatrix}
\ddot{y}_i \\ \addlinespace \ddot{x}_i \\ \addlinespace 0
\end{pmatrix} + \frac{2}{r_{i_{xy}}^2}\begin{pmatrix}
-2x_i \dot{x}_i \dot{y}_i - y \dot{y}_i^2 + y \dot{x}_i^2 \\ \addlinespace 2y_i \dot{x}_i \dot{y}_i + x \dot{x}_i^2 - x \dot{y}_i^2
\\ \addlinespace 0
\end{pmatrix} \nonumber \\
&+\frac{4\left(x_i\dot{y}_i - y_i \dot{x}_i \right)\left(x_i \dot{x}_i + y_i \dot{y}_i\right)}{r_{i_{xy}}^4}\begin{pmatrix}
x \\ y \\ 0
\end{pmatrix}.
\end{align}
The partials of $\ddot{\alpha}_{_{Si}}$ with respect to the position vectors of the observer and of the other satellite $j$ are
\begin{align}
\left.\frac{\partial \ddot{\alpha}_{_{Si}}}{\partial \boldsymbol{r}_{_{O}}}\right|_{t_{_{Si}}} &= \begin{pmatrix}
x_i \frac{\partial \ddot{y}_i}{\partial x_{_O}} - y_i \frac{\partial \ddot{x}_i}{\partial x_{_O}} \\ \addlinespace
x_i \frac{\partial \ddot{y}_i}{\partial y_{_O}} - y_i \frac{\partial \ddot{x}_i}{\partial y_{_O}}
\\ \addlinespace x_i \frac{\partial \ddot{y}_i}{\partial z_{_O}} - y_i \frac{\partial \ddot{x}_i}{\partial z_{_O}}
\end{pmatrix} - g^{\prime}_{\ddot{\alpha}_{_{Si}}}, 
\end{align}
\begin{align}
\left.\frac{\partial \ddot{\alpha}_{_{Si}}}{\partial \boldsymbol{r}_{_{Sj}}}\right|_{t_{_{Si}}} &= \begin{pmatrix}
x_i \frac{\partial \ddot{y}_i}{\partial x_{_{Sj}}} - y_i \frac{\partial \ddot{x}_i}{\partial x_{_{Sj}}} \\ \addlinespace
x_i \frac{\partial \ddot{y}_i}{\partial y_{_{Sj}}} - y_i \frac{\partial \ddot{x}_i}{\partial y_{_{Sj}}}
\\ \addlinespace x_i \frac{\partial \ddot{y}_i}{\partial z_{_{Sj}}} - y_i \frac{\partial \ddot{x}_i}{\partial z_{_{Sj}}}
\end{pmatrix} ; i \neq j.
\end{align}
Similarly, the position partials of $\ddot{\delta}_{_{Si}}$ are written as follows:
\begin{align}
\left.\frac{\partial \ddot{\delta}_{_{Si}}}{\partial \boldsymbol{r}_{_{Si}}}\right|&_{t_{_{Si}}} = g_{\ddot{\delta}_{_{Si}}} + g^{\prime}_{\ddot{\delta}_{_{Si}}}, \text{ with}
\end{align}
\begin{align}
g_{\ddot{\delta}_{_{Si}}} &= \frac{-z_i }{r_{i}^2r_{i_{xy}}}\begin{pmatrix}
x_i \frac{\partial \ddot{x}_i}{\partial {x}_{_{Si}}}
+y_i \frac{\partial \ddot{y}_i}{\partial {x}_{_{Si}}}\\ \addlinespace
x_i \frac{\partial \ddot{x}_i}{\partial {y}_{_{Si}}}
+y_i \frac{\partial \ddot{y}_i}{\partial {y}_{_{Si}}}\\ \addlinespace
x_i \frac{\partial \ddot{x}_i}{\partial {z}_{_{Si}}}
+y_i \frac{\partial \ddot{y}_i}{\partial {z}_{_{Si}}}
\end{pmatrix} +\frac{r_{i_{xy}}}{r_{i}^2}\begin{pmatrix}
\frac{\partial \ddot{z}_i}{\partial {x}_{_{Si}}}\\ \addlinespace
\frac{\partial \ddot{z}_i}{\partial {y}_{_{Si}}}\\ \addlinespace
\frac{\partial \ddot{z}_i}{\partial {z}_{_{Si}}}
\end{pmatrix}, \\
g^{\prime}_{\ddot{\delta}_{_{Si}}}
&= \frac{1}{r_{i}^2r_{i_{xy}}}\begin{pmatrix}
-z_i \ddot{x}_i +2 x_i \ddot{z}_i \\ \addlinespace
-z_i \ddot{y}_i +2 y_i \ddot{z}_i\\ \addlinespace
-x_i \ddot{x}_i -y_i \ddot{y}_i
\end{pmatrix} -\frac{\ddot{\delta}_{_{Si}}}{r_{i}^2r_{i_{xy}}^2}\begin{pmatrix}
x_i(r_{i_{xy}}^2-z_i^2)\\ \addlinespace
y_i(r_{i_{xy}}^2-z_i^2)\\ \addlinespace
2 z_i r_{i_{xy}}^2
\end{pmatrix} \nonumber \\
&-4\left(z_i\left(x_i \dot{x}_i + y_i \dot{y}_i\right)-\dot{z}_ir_{i_{xy}}^2\right)\left(\boldsymbol{r}_{_O}^{S_{i}}\cdot \dot{\boldsymbol{r}}_{_O}^{S_{i}}\right)\boldsymbol{r}_{_O}^{S_{i}} \nonumber \\
&-\frac{x_i \dot{y}_i-y_i \dot{x}_i}{r_{i}^2r_{i_{xy}}^5}\begin{pmatrix}
2z_i\left(y_i^2\dot{y}_i+x_i y_i \dot{x}_i\right)\\ \addlinespace
2z_i\left(x_i^2\dot{x}_i+2y_i^2 \dot{x}_i - x_i y_i \dot{y}_i\right) \\ \addlinespace
-r_{i_{xy}}^2\left(x_i \dot{y}_i - y_i \dot{x}_i\right)
\end{pmatrix} \nonumber \\
&+\frac{2}{r_{i}^2}\begin{pmatrix}
-2x_i\dot{z}_i\left(y_i \dot{y}_i + z_i \dot{z}_i\right) + \dot{x}_i \dot{z}_i\left(-3x_i^2 -y_i^2 + z_i^2\right) \\ \addlinespace
-2y_i\dot{z}_i\left(x_i \dot{x}_i + z_i \dot{z}_i\right) + \dot{y}_i \dot{z}_i\left(-x_i^2 -3y_i^2 + z_i^2\right)\\ \addlinespace
-r_{i_{xy}}^2\dot{z}_i^2
\end{pmatrix}\nonumber \\
&+\frac{2\left(x \dot{x}_i + y_i \dot{y}_i\right)}{r_{i}^2}\begin{pmatrix}
2z_i \dot{x}_i \\ \addlinespace
2z_i \dot{y}_i \\ \addlinespace
2z_i \dot{z}_i+x_i \dot{x}_i+y_i \dot{y}_i
\end{pmatrix}, 	\\ 
&\nonumber\\
\left.\frac{\partial \ddot{\delta}_{_{Si}}}{\partial \boldsymbol{r}_{_{O}}}\right|_{t_{_{Si}}} &= \frac{-z_i }{r_{i}^2r_{i_{xy}}}\begin{pmatrix}
x_i \frac{\partial \ddot{x}_i}{\partial {x}_{_{O}}}
+y_i \frac{\partial \ddot{y}_i}{\partial {x}_{_{O}}}\\ \addlinespace
x_i \frac{\partial \ddot{x}_i}{\partial {y}_{_{O}}}
+y_i \frac{\partial \ddot{y}_i}{\partial {y}_{_{O}}}\\ \addlinespace
x_i \frac{\partial \ddot{x}_i}{\partial {z}_{_{O}}}
+y_i \frac{\partial \ddot{y}_i}{\partial {z}_{_{O}}}
\end{pmatrix} +\frac{r_{i_{xy}}}{r_{i}^2}\begin{pmatrix}
\frac{\partial \ddot{z}_i}{\partial {x}_{_{O}}}\\ \addlinespace
\frac{\partial \ddot{z}_i}{\partial {y}_{_{O}}}\\ \addlinespace
\frac{\partial \ddot{z}_i}{\partial {z}_{_{O}}}
\end{pmatrix} - g^{\prime}_{\ddot{\delta}_{_{Si}}}, 
\end{align}
\begin{align}	
\left.\frac{\partial \ddot{\delta}_{_{Si}}}{\partial \boldsymbol{r}_{_{Sj}}}\right|_{t_{_{Si}}} &= \frac{-z_i }{r_{i}^2r_{i_{xy}}}\begin{pmatrix}
x_i \frac{\partial \ddot{x}_i}{\partial {x}_{_{Sj}}}
+y_i \frac{\partial \ddot{y}_i}{\partial {x}_{_{Sj}}}\\ \addlinespace
x_i \frac{\partial \ddot{x}_i}{\partial {y}_{_{Sj}}}
+y_i \frac{\partial \ddot{y}_i}{\partial {y}_{_{Sj}}}\\ \addlinespace
x_i \frac{\partial \ddot{x}_i}{\partial {z}_{_{Sj}}}
+y_i \frac{\partial \ddot{y}_i}{\partial {z}_{_{Sj}}}
\end{pmatrix} +\frac{r_{i_{xy}}}{r_{i}^2}\begin{pmatrix}
\frac{\partial \ddot{z}_i}{\partial {x}_{_{S_j}}}\\ \addlinespace
\frac{\partial \ddot{z}_i}{\partial {y}_{_{S_j}}}\\ \addlinespace
\frac{\partial \ddot{z}_i}{\partial {z}_{_{S_j}}}
\end{pmatrix} ; i \neq j.
\end{align}

Finally, the velocity partials also have to be derived for $\ddot{\alpha}_{_{Si}}$ and $\ddot{\delta}_{_{Si}}$. We again split the partials expressions into two terms, designated by $k_{\ddot{\alpha}_{_{Si}}}$ and $k^{\prime}_{\ddot{\alpha}_{_{Si}}}$ (or $k_{\ddot{\delta}_{_{Si}}}$ and $k^{\prime}_{\ddot{\delta}_{_{Si}}}$). Again $k_{\ddot{\alpha}_{_{Si}}}$ corresponds to the contribution of the acceleration partials. Starting with the partials of $\ddot{\alpha}_{_{Si}}$ with respect to the satellites' and observer's velocity vectors, we obtain the following expressions: 
\begin{align}
\left.\frac{\partial \ddot{\alpha}_{_{Si}}}{\partial \dot{\boldsymbol{r}}_{_{Si}}}\right|_{t_{_{Si}}} &=k_{\ddot{\alpha}_{_{Si}}} + k^{\prime}_{\ddot{\alpha}_{_{Si}}}, \text{ with} \\
k_{\ddot{\alpha}_{_{Si}}} &= \frac{1}{r_{i_{xy}}^2}\begin{pmatrix}
x_i \frac{\partial \ddot{y}_i}{\partial \dot{x}_{_{Si}}} - y_i \frac{\partial \ddot{x}_i}{\partial \dot{x}_{_{Si}}}\\ \addlinespace x_i \frac{\partial \ddot{y}_i}{\partial \dot{y}_{_{Si}}} - y_i \frac{\partial \ddot{x}_i}{\partial \dot{y}_{_{Si}}} \\ \addlinespace x_i \frac{\partial \ddot{y}_i}{\partial \dot{z}_{_{Si}}} - y_i \frac{\partial \ddot{x}_i}{\partial \dot{z}_{_{Si}}}
\end{pmatrix},  \\
k^{\prime}_{\ddot{\alpha}_{_{Si}}} &= \frac{2}{r_{i_{xy}}^4} \begin{pmatrix}
(y_i^2 - x_i^2)\dot{y}_i + 2x_i y_i \dot{x}_{i}\\ \addlinespace (y_i^2 - x_i^2)\dot{x}_i - 2x_i y_i \dot{y}_{i} \\ \addlinespace 0
\end{pmatrix}, \\
\left.\frac{\partial \ddot{\alpha}_{_{Si}}}{\partial \dot{\boldsymbol{r}}_{_{O}}}\right|_{t_{_{Si}}} &= \frac{1}{r_{i_{xy}}^2}\begin{pmatrix}
x_i \frac{\partial \ddot{y}_i}{\partial \dot{x}_{_{O}}} - y_i \frac{\partial \ddot{x}_i}{\partial \dot{x}_{_{O}}}\\ \addlinespace x_i \frac{\partial \ddot{y}_i}{\partial \dot{y}_{_{O}}} - y_i \frac{\partial \ddot{x}_i}{\partial \dot{y}_{_{O}}} \\ \addlinespace x_i \frac{\partial \ddot{y}_i}{\partial \dot{z}_{_{O}}} - y_i \frac{\partial \ddot{x}_i}{\partial \dot{z}_{_{O}}}
\end{pmatrix}- k^{\prime}_{\ddot{\alpha}_{_{Si}}}, \\
\left.\frac{\partial \ddot{\alpha}_{_{Si}}}{\partial \dot{\boldsymbol{r}}_{_{Sj}}}\right|_{t_{_{Si}}} &= \frac{1}{r_{i_{xy}}^2}\begin{pmatrix}
x_i \frac{\partial \ddot{y}_i}{\partial \dot{x}_{_{Sj}}} - y_i \frac{\partial \ddot{x}_i}{\partial \dot{x}_{_{Sj}}}\\ \addlinespace x_i \frac{\partial \ddot{y}_i}{\partial \dot{y}_{_{Sj}}} - y_i \frac{\partial \ddot{x}_i}{\partial \dot{y}_{_{Sj}}} \\ \addlinespace x_i \frac{\partial \ddot{y}_i}{\partial \dot{z}_{_{Sj}}} - y_i \frac{\partial \ddot{x}_i}{\partial \dot{z}_{_{Sj}}}
\end{pmatrix}; i \neq j.
\end{align}
Lastly, we obtain the following formulations for the partials of $\ddot{\delta}_{_{Si}}$ with respect to the satellites' and observer's velocity vectors: 
\begin{align}
\left.\frac{\partial \ddot{\delta}_{_{Si}}}{\partial \dot{\boldsymbol{r}}_{_{Si}}}\right|_{t_{_{Si}}} &=k_{\ddot{\delta}_{_{Si}}} + k^{\prime}_{\ddot{\delta}_{_{Si}}}, \text{ with} \\
k_{\ddot{\delta}_{_{Si}}} &= \frac{-z_i}{r_{i}^2r_{i_{xy}}} \begin{pmatrix} x_i  \frac{\partial \ddot{x}_i}{\partial \dot{x}_{_{Si}}}
+y_i \frac{\partial \ddot{y}_i}{\partial \dot{x}_{_{Si}}}\\ \addlinespace x_i \frac{\partial \ddot{x}_i}{\partial \dot{y}_{_{Si}}}
+y_i \frac{\partial \ddot{y}_i}{\partial \dot{y}_{_{Si}}} \\ \addlinespace x_i \frac{\partial \ddot{x}_i}{\partial \dot{z}_{_{Si}}}
+ y_i \frac{\partial \ddot{y}_i}{\partial \dot{z}_{_{Si}}}  \end{pmatrix} +\frac{r_{i_{xy}}}{r_{i}^2} \begin{pmatrix}  \frac{\partial \ddot{z}_i}{\partial \dot{x}_{_{Si}}}\\ \addlinespace \frac{\partial \ddot{z}_i}{\partial \dot{y}_{_{Si}}} \\ \addlinespace \frac{\partial \ddot{z}_i}{\partial \dot{z}_{_{Si}}} \end{pmatrix}, \\
k^{\prime}_{\ddot{\delta}_{_{Si}}} &= \frac{2 z_i \left(y_i \dot{x}_i - x_i \dot{y}_i\right)}{r_{i_{xy}}^2} \begin{pmatrix} -y_i \\ x_i \\ 0 \end{pmatrix} \nonumber \\
&+\frac{2}{r_{_i}^2}\begin{pmatrix}
- x_i \dot{z}_i\left(r_{i_{xy}}^2 - z_i^2\right) + 2 x_i z_i \left(x_i \dot{x}_i + y_i \dot{y}_i\right) \\ \addlinespace - y_i \dot{z}_i\left(r_{i_{xy}}^2 - z_i^2\right) + 2 y_i z_i \left(x_i \dot{x}_i + y_i \dot{y}_i\right) \\ \addlinespace - \left(r_{i_{xy}}^2 - z_i^2\right)\left(x_i \dot{x}_i + y_i \dot{y}_i\right) - 2 z_i \dot{z}_i r_{i_{xy}}^2
\end{pmatrix}, \\
\left.\frac{\partial \ddot{\delta}_{_{Si}}}{\partial \dot{\boldsymbol{r}}_{_{O}}}\right|_{t_{_{Si}}} &= \frac{-z_i}{r_{i}^2r_{i_{xy}}} \begin{pmatrix} x_i  \frac{\partial \ddot{x}_i}{\partial \dot{x}_{_{O}}}
+y_i \frac{\partial \ddot{y}_i}{\partial \dot{x}_{_{O}}}\\ \addlinespace x_i \frac{\partial \ddot{x}_i}{\partial \dot{y}_{_{O}}}
+y_i \frac{\partial \ddot{y}_i}{\partial \dot{y}_{_{O}}} \\ \addlinespace x_i \frac{\partial \ddot{x}_i}{\partial \dot{z}_{_{O}}}
+ y_i \frac{\partial \ddot{y}_i}{\partial \dot{z}_{_{O}}}  \end{pmatrix} +\frac{r_{i_{xy}}}{r_{i}^2} \begin{pmatrix}  \frac{\partial \ddot{z}_i}{\partial \dot{x}_{_{O}}}\\ \addlinespace \frac{\partial \ddot{z}_i}{\partial \dot{y}_{_{O}}} \\ \addlinespace  \frac{\partial \ddot{z}_i}{\partial \dot{z}_{_{O}}} \end{pmatrix} - k^{\prime}_{\ddot{\delta}_{_{Si}}},
\end{align}
\begin{align}
\left.\frac{\partial \ddot{\delta}_{_{Si}}}{\partial \dot{\boldsymbol{r}}_{_{Sj}}}\right|_{t_{_{Si}}} &= \frac{-z_i}{r_{i}^2r_{i_{xy}}} \begin{pmatrix} x_i  \frac{\partial \ddot{x}_i}{\partial \dot{x}_{_{Sj}}}
+y_i \frac{\partial \ddot{y}_i}{\partial \dot{x}_{_{Sj}}}\\ \addlinespace x_i \frac{\partial \ddot{x}_i}{\partial \dot{y}_{_{Sj}}}
+y_i \frac{\partial \ddot{y}_i}{\partial \dot{y}_{_{Sj}}} \\ \addlinespace x_i \frac{\partial \ddot{x}_i}{\partial \dot{z}_{_{Sj}}}
+ y_i \frac{\partial \ddot{y}_i}{\partial \dot{z}_{_{Sj}}}  \end{pmatrix} +\frac{r_{i_{xy}}}{r_{i}^2} \begin{pmatrix}  \frac{\partial \ddot{z}_i}{\partial \dot{x}_{_{Sj}}}\\ \addlinespace \frac{\partial \ddot{z}_i}{\partial \dot{y}_{_{Sj}}} \\ \addlinespace \frac{\partial \ddot{z}_i}{\partial \dot{z}_{_{Sj}}} \end{pmatrix}; i \neq j. 
\end{align}

\section{Acceleration partials} \label{appendix:accelerationPartials}
As shown in Appendix \ref{appendix:partials}, computing $\ddot{\alpha}_{_{S_{i}}}$ and $\ddot{\delta}_{_{S_{i}}}$  partials requires to first compute the partials of those relative acceleration, starting from Equation \ref{eqn:inertialRelativeAcceleration}:
\begin{align}
\frac{\partial \ddot{\boldsymbol{r}}_{_O}^{S_i}}{\partial \boldsymbol{q}} & = \frac{\partial \ddot{\boldsymbol{r}}_{_{Si}}(t_{_{Si}})}{\partial \boldsymbol{q}} - \frac{\partial\ddot{\boldsymbol{r}}_{_O}(t_{_O})}{\partial \boldsymbol{q}} ; \text{ } i\in \{1,2\}.
\end{align}
The vector of parameters $\boldsymbol{q}$ can either refer to one of the satellites state $\boldsymbol{s}_{_{Si}}(t_{_{Si}})$ or to the observer state $\boldsymbol{s}_{_O}(t_{_O})$. We first consider the partials with respect to the observer state, given by
\begin{align}
\frac{\partial \ddot{\boldsymbol{r}}_{_O}^{S_i}}{\partial \boldsymbol{s}_{_O}(t_{_O})} & = \frac{\partial \ddot{\boldsymbol{r}}_{_{Si}}(t_{_{Si}})}{\partial \boldsymbol{s}_{_O}(t_{_O})} - \frac{\partial\ddot{\boldsymbol{r}}_{_O}(t_{_O})}{\partial\boldsymbol{s}_{_O}(t_{_O})} ; \text{ } i\in \{1,2\}. \label{eqn:accelerationPartialWrtReceiverState1}
\end{align}

The acceleration $\ddot{\boldsymbol{r}}_{_{Si}}(t_{_{Si}})$ of the satellite $i$ at time $t_{_{Si}}$ depends on the observer state $\boldsymbol{s}_{_O}$ at $t = t_{_{Si}}$, assuming the observer's body indeed exerts an acceleration on satellite $i$ (although such acceleration is usually negligible, see simplifying assumptions discussed at the end of this appendix). Equation \ref{eqn:accelerationPartialWrtReceiverState1} must thus be rewritten as
\begin{align}
\frac{\partial \ddot{\boldsymbol{r}}_{_O}^{S_i}}{\partial \boldsymbol{s}_{_O}(t_{_O})} &= \frac{\partial \ddot{\boldsymbol{r}}_{_{Si}}(t_{_{Si}})}{\partial \boldsymbol{s}_{_O}(t_{_{Si}})} \frac{\boldsymbol{s}_{_O}(t_{_{Si}})}{\boldsymbol{s}_{_O}(t_{_{O}})} - \frac{\partial\ddot{\boldsymbol{r}}_{_O}(t_{_O})}{\partial\boldsymbol{s}_{_O}(t_{_O})} \label{eqn:accelerationPartialWrtReceiverState2} \nonumber \\ 
&= \frac{\partial \ddot{\boldsymbol{r}}_{_{Si}}(t_{_{Si}})}{\partial \boldsymbol{s}_{_O}(t_{_{Si}})}\boldsymbol{\Phi}_{_O}(t_{_O}, t_{_{Si}}) - \frac{\partial\ddot{\boldsymbol{r}}_{_O}(t_{_O})}{\partial\boldsymbol{s}_{_O}(t_{_O})} ; \text{ } i\in \{1,2\}. 
\end{align}
Similarly, acceleration partials with respect to the two satellites' states are expressed as follows:
\begin{align}
\frac{\partial \ddot{\boldsymbol{r}}_{_O}^{S_i}}{\partial \boldsymbol{s}_{_{Si}}(t_{_{Si}})} &= \frac{\partial \ddot{\boldsymbol{r}}_{_{Si}}(t_{_{Si}})}{\partial \boldsymbol{s}_{_{Si}}(t_{_{Si}})} - \frac{\partial\ddot{\boldsymbol{r}}_{_O}(t_{_O})}{\partial\boldsymbol{s}_{_{Si}}(t_{_O})}\boldsymbol{\Phi}_{_{Si}}(t_{_{Si}},t_{_O}), \label{eqn:accelerationPartialWrtTransmitterI} \\
\frac{\partial \ddot{\boldsymbol{r}}_{_O}^{S_i}}{\partial \boldsymbol{s}_{_{Sj}}(t_{_{Sj}})} &= \frac{\partial \ddot{\boldsymbol{r}}_{_{Si}}(t_{_{Si}})}{\partial \boldsymbol{s}_{_{Sj}}(t_{_{Si}})}\boldsymbol{\Phi}_{_{Sj}}(t_{_{Sj}},t_{_{Si}}) \label{eqn:accelerationPartialWrtTransmitterJ} \nonumber \\
&- \frac{\partial\ddot{\boldsymbol{r}}_{_O}(t_{_O})}{\partial\boldsymbol{s}_{_{Sj}}(t_{_O})}\boldsymbol{\Phi}_{_{Sj}}(t_{_{Sj}},t_{_O}) ; \text{ } \{i,j\} \in \{1,2\},  j \neq i. 
\end{align}

According to Equations \ref{eqn:accelerationPartialWrtReceiverState2}-\ref{eqn:accelerationPartialWrtTransmitterJ}, four state transition matrices need to be computed. However, a few remarks must be considered, in light of the computational effort this would require. For mutual approximations between the Galilean moons observed from the Earth, the satellite-observer distance is comparable between the two satellites. The difference between the two times $t_{_{S1}}$ and $t_{_{S2}}$ is thus very small, and the state transition matrices $\boldsymbol{\Phi}_{_{Si}}(t_{_{Si}},t_{_{Sj}})$ (with $\{i,j\} \in \{1,2\}$ and $j \neq i$) are consequently close to unit matrices. 

The difference between each time $t_{_{Si}}$ and the observation time $t_{_O}$ is larger. However, looking at Equations \ref{eqn:accelerationPartialWrtReceiverState2}-\ref{eqn:accelerationPartialWrtTransmitterJ}, the state transition matrices $\boldsymbol{\Phi}_{_O}(t_{_O}, t_{_{Si}})$ and $\boldsymbol{\Phi}_{_{Si}}(t_{_{Si}},t_{_O})$ are always multiplied by partials of the observer's body acceleration with respect to one of the satellite's state, or the other way around.  Considering the large satellites - observer distances, these accelerations partials can actually be neglected. 

Overall, for mutual approximations between Galilean moons, the state transition matrices appearing in Equations \ref{eqn:accelerationPartialWrtReceiverState2}-\ref{eqn:accelerationPartialWrtTransmitterJ} can be either approximated by unit matrices, or the entire acceleration partial term they contribute to can be neglected. This significantly simplifies the implementation and reduces the computational effort. Acceleration partials are anyway only required to compute the partials of $\ddot{\alpha}_{_{S_{i}}}$ and $\ddot{\delta}_{_{S_{i}}}$, which represent a marginal contribution of the total central instant partials (see Section \ref{sec:lightTimeCorrections}). Simplifying assumptions to compute those acceleration partials can therefore be made safely. 

\section{Verification of the analytical partials} \label{appendix:validationPartials}

\setcounter{table}{0}

The central instants partials derived in Section \ref{sec:centralInstantPartials} were validated numerically, by comparing the analytically estimated change in central instant with the actual change obtained when applying a small variation to the estimated parameters. Partials were expressed with respect to $\boldsymbol{r}_{_{S1}}(t_{_{S1}})$, $\boldsymbol{r}_{_{S2}}(t_{_{S2}})$ and $\boldsymbol{r}_{_{O}}(t_{_{O}})$ (for the first satellite's, second satellite's and observer's states, respectively). 
Analytical approximations of the changes in central instants were derived from the observation partials with respect to the initial state of interest, multiplied with the appropriate state transition matrix, as follows:
\begin{align}
\Delta t_c &= \frac{\partial t_c}{\partial \boldsymbol{r}_{_{S1}}(t_{_{S1}})}\boldsymbol{\Phi}\left(t_{_{S1}}, t_0\right)\Delta \boldsymbol{r}_{_{S1}}(t_{0}), \\
\Delta t_c &= \frac{\partial t_c}{\partial \boldsymbol{r}_{_{S2}}(t_{_{S2}})}\boldsymbol{\Phi}\left(t_{_{S2}}, t_0\right)\Delta \boldsymbol{r}_{_{S2}}(t_{0}),\\
\Delta t_c &= \frac{\partial t_c}{\partial \boldsymbol{r}_{_{O}}(t_{_{O}})}\boldsymbol{\Phi}\left(t_{_{O}}, t_0\right)\Delta \boldsymbol{r}_{_{O}}(t_{0}).
\end{align}

The results of the numerical verification are reported in Table \ref{tab:validationPartials}. The extremely low differences found between the analytical and numerical changes in central instant prove the validity of our analytical partials. 

\begin{table}[h]
	\caption{Comparison between analytical and numerical solutions for the changes in central instants.
	}
	\label{tab:validationPartials}
	\centering
	\begin{tabular}{c c c c}
		\hline \hline \noalign{\vskip 0.02in}		
		Mutual & \multicolumn{2}{c}{Change in $t_c$} & Relative \\
		approx. & analytical [s] & numerical [s] & error [-] \\ \hline \noalign{\vskip 0.02in}
		1&2.17458&2.17455&$7.84\cdot 10^{-6}$\\
		2&36.7842&36.7824&$4.85\cdot 10^{-5}$\\
		3&73.9007&73.8972&$4.76\cdot 10^{-5}$\\
		4&94.9994&94.9951&$4.56\cdot 10^{-5}$\\
		5&111.043&111.038&$4.74\cdot 10^{-5}$\\
		6&132.098&132.092&$4.55\cdot 10^{-5}$\\
		7&169.169&169.161&$4.55\cdot 10^{-5}$\\
		8&203.983&203.974&$4.58\cdot 10^{-5}$\\
		9&206.006&206.198&$9.30\cdot 10^{-4}$\\
		10&241.156&241.145&$4.53\cdot 10^{-5}$\\
		11&261.643&261.631&$4.35\cdot 10^{-5}$\\
		12&278.322&278.309&$4.51\cdot 10^{-5}$\\
		13&298.557&298.544&$4.21\cdot 10^{-5}$\\
		14&335.416&335.403&$4.04\cdot 10^{-5}$\\
		15&371.133&371.117&$4.27\cdot 10^{-5}$\\
		16&372.219&372.205&$3.84\cdot 10^{-5}$\\
		17&408.219&408.202&$4.23\cdot 10^{-5}$\\
		18&445.273&445.254&$4.21\cdot 10^{-5}$\\
		19&464.009&463.996&$2.82\cdot 10^{-5}$\\
		20&482.288&482.267&$4.21\cdot 10^{-5}$\\ \hline                                                                               
	\end{tabular}
	\tablefoot{This table compares the analytical and numerical solutions for the changes in central instants after applying a small variation (0.001\%) in the initial states of Io, Europa and the Earth. Analytical approximations of the changes are derived from the central instants partials provided in Section \ref{sec:centralInstantPartials}. Results are here only reported for the 20 first mutual approximations detected in 2020 (although verification was conducted over 200 observations).}
\end{table}

\section{Contribution of the $\ddot{\alpha}_{_{Si}}$ and $\ddot{\delta}_{_{Si}}$ partials to the central instant partials} \label{appendix:LTcontributions}

\setcounter{table}{0}

Table \ref{tab:contributionPartialsSecondTimeDerivativeRightAscensionDeclination} gives the relative contributions of the $\ddot{\alpha}_{_{Si}}$ and $\ddot{\delta}_{_{Si}}$ partials to the total central instant partials. They are reported for the first five Io-Europa mutual approximations in 2020 and are shown to be negligible. 

\begin{table*}[h]
	\caption{Relative contributions of the $\ddot{\alpha}_{_{Si}}$ and $\ddot{\delta}_{_{Si}}$ partials to the total central instants partials.}
	\label{tab:contributionPartialsSecondTimeDerivativeRightAscensionDeclination}
	\centering
	\begin{tabular}{c c c c c c c}
		\hline \hline \noalign{\vskip 0.02in}
		Mutual & \multicolumn{6}{c}{Relative contribution to the $t_c$ partials [\%]} \\
		approx. &  & &  &  &  & \\ \hline \noalign{\vskip 0.03in}
		& \multicolumn{6}{c}{w.r.t. first satellite's state $\boldsymbol{s}_{_{S1}} = [\boldsymbol{r}_{_{S1}} \text{ } \dot{\boldsymbol{r}}_{_{S1}}]^{\mathrm{T}}$} \\ 
		& $x_{_{S1}}$ & $y_{_{S1}}$ & $z_{_{S1}}$ & $\dot{x}_{_{S1}}$ & $\dot{y}_{_{S1}}$ & $\dot{z}_{_{S1}}$ \\ 
		\noalign{\vskip 0.03in}
		1&$6.8\cdot10^{-7}$&$1.3\cdot10^{-4}$&$3.2\cdot10^{-4}$&$1.1\cdot10^{-5}$&$9.1\cdot10^{-8}$&$9.3\cdot10^{-8}$\\ 
		2&$5.4\cdot10^{-6}$&$2.8\cdot10^{-4}$&$1.0\cdot10^{-3}$&$1.3\cdot10^{-4}$&$1.0\cdot10^{-6}$&$6.4\cdot10^{-7}$\\ 
		3&$4.9\cdot10^{-6}$&$2.2\cdot10^{-4}$&$8.0\cdot10^{-4}$&$1.1\cdot10^{-4}$&$8.8\cdot10^{-7}$&$3.9\cdot10^{-7}$\\ 
		4&$1.2\cdot10^{-6}$&$9.8\cdot10^{-5}$&$2.9\cdot10^{-4}$&$1.3\cdot10^{-5}$&$1.4\cdot10^{-8}$&$4.7\cdot10^{-8}$\\ 
		5&$3.0\cdot10^{-6}$&$1.2\cdot10^{-5}$&$4.4\cdot10^{-4}$&$6.5\cdot10^{-5}$&$5.1\cdot10^{-7}$&$1.0\cdot10^{-7}$\\ \noalign{\vskip 0.02in} \hline \noalign{\vskip 0.03in}
		& \multicolumn{6}{c}{w.r.t. second satellite's state $\boldsymbol{s}_{_{S2}} = [\boldsymbol{r}_{_{S2}} \text{ } \dot{\boldsymbol{r}}_{_{S2}}]^{\mathrm{T}}$} \\ 
		& $x_{_{S2}}$ & $y_{_{S2}}$ & $z_{_{S2}}$ & $\dot{x}_{_{S2}}$ & $\dot{y}_{_{S2}}$ & $\dot{z}_{_{S2}}$ \\ \noalign{\vskip 0.03in} 
		1&$2.5\cdot10^{-7}$&$3.3\cdot10^{-5}$&$7.9\cdot10^{-5}$&$2.2\cdot10^{-5}$&$8.7\cdot10^{-8}$&$1.2\cdot10^{-9}$\\
		2&$7.1\cdot10^{-7}$&$7.1\cdot10^{-5}$&$2.6\cdot10^{-4}$&$7.8\cdot10^{-5}$&$2.7\cdot10^{-7}$&$1.6\cdot10^{-7}$\\
		3&$6.5\cdot10^{-7}$&$5.5\cdot10^{-5}$&$2.0\cdot10^{-4}$&$6.4\cdot10^{-5}$&$1.4\cdot10^{-7}$&$4.7\cdot10^{-8}$\\
		4&$3.5\cdot10^{-7}$&$2.6\cdot10^{-5}$&$7.2\cdot10^{-5}$&$2.6\cdot10^{-5}$&$1.0\cdot10^{-8}$&$2.2\cdot10^{-7}$\\
		5&$4.1\cdot10^{-7}$&$3.1\cdot10^{-5}$&$1.1\cdot10^{-4}$&$3.7\cdot10^{-5}$&$3.4\cdot10^{-9}$&$1.7\cdot10^{-7}$\\ \noalign{\vskip 0.02in} \hline \noalign{\vskip 0.03in}
		& \multicolumn{6}{c}{w.r.t. observer's state $\boldsymbol{s}_{_{O}} = [\boldsymbol{r}_{_{O}} \text{ } \dot{\boldsymbol{r}}_{_{O}}]^{\mathrm{T}}$} \\
		& $x_{_{O}}$ & $y_{_{O}}$ & $z_{_{O}}$ & $\dot{x}_{_{O}}$ & $\dot{y}_{_{O}}$ & $\dot{z}_{_{O}}$ \\  \noalign{\vskip 0.03in}
		1&$1.7\cdot10^{-6}$&$5.3\cdot10^{-5}$&$1.7\cdot10^{-4}$&$8.6\cdot10^{-3}$&$3.5\cdot10^{-6}$&$8.3\cdot10^{-5}$\\
		2&$1.0\cdot10^{-5}$&$9.0\cdot10^{-5}$&$6.1\cdot10^{-4}$&$4.4\cdot10^{-2}$&$9.2\cdot10^{-4}$&$5.5\cdot10^{-4}$\\
		3&$8.7\cdot10^{-6}$&$7.2\cdot10^{-5}$&$4.6\cdot10^{-4}$&$3.7\cdot10^{-2}$&$9.0\cdot10^{-4}$&$4.8\cdot10^{-4}$\\
		4&$2.8\cdot10^{-6}$&$4.3\cdot10^{-5}$&$1.5\cdot10^{-4}$&$9.6\cdot10^{-3}$&$2.4\cdot10^{-5}$&$1.5\cdot10^{-4}$\\
		5&$5.1\cdot10^{-6}$&$4.1\cdot10^{-5}$&$2.5\cdot10^{-4}$&$2.2\cdot10^{-2}$&$6.0\cdot10^{-4}$&$2.9\cdot10^{-4}$\\
		\hline                            
	\end{tabular}
	\tablefoot{The partials are computed with respect to the first satellite's state $\boldsymbol{s}_{_{S1}}$, second satellite's state $\boldsymbol{s}_{_{S2}}$ and observer's state $\boldsymbol{s}_{_{O}}$, all expressed in cartesian coordinates. Results are only reported for 5 mutual approximations (five first Io-Europa mutual approximations from 01/01/2020).}
\end{table*}

\section{Alternative observables' weights for past mutual approximations (2016-2018 observational campaign)} \label{appendix:weightsPastObservations}

We computed the alternative observables' weights for the past mutual approximations observed during the 2016-2018 campaign, which are provided in \cite{morgado2019approx}. Tables \ref{tab:weightsPastObservations2016}-\ref{tab:weightsPastObservations2018} contain the weight values obtained with Equation \ref{eqn:alternativeObservableWeight}, following the weighting strategy described in Section \ref{section:dataWeights} (we have shown this approach to be crucial to obtain consistent estimation solutions between central instants and alternative observable in Section \ref{sec:influenceWeightingScheme}). 

For consistency purposes, the errors in central instant $\sigma(t_c)$ given in \cite{morgado2019approx} were translated into errors in the alternative observable $\sigma_{\mathrm{alt.}}$ using the same ephemerides as the ones used in \cite{morgado2019approx} (i.e. JUP310 with DE435 from JPL). Table \ref{tab:weightsPastObservations2016} corresponds to mutual approximations observed in 2016, while Tables \ref{tab:weightsPastObservations2017} and \ref{tab:weightsPastObservations2018} are dedicated to 2017 and 2018 observations, respectively. The geodetic coordinates of the six different ground stations can be found in \cite{morgado2019approx}. We recommend to use the computed weight values $\sigma_{\mathrm{alt.}}$ when including the 2016-2018 observations in the state estimation with alternative observables. 

\begin{table*}[h]
	\caption{Appropriate weights for alternative observables, for the mutual approximations observed in 2016.}
	\label{tab:weightsPastObservations2016}
	\centering
	\begin{tabular}{cccccc} \hline \hline \noalign{\vskip 0.03in}
		Date & Event & Station & $t_c$ (UTC) & $\sigma(t_c)$ [s] & $\sigma_{\mathrm{alt.}}$ [mas/s]\\ \hline \noalign{\vskip 0.03in}
		2016-02-03 & E-G & OPD & 04:48:01.1 & 4.2  & $1.357\cdot 10^{-2}$ \\
		2016-02-08 & I-E & FOZ & 06:29:38.4 & 0.6  & $2.428\cdot 10^{-3}$ \\
		2016-02-15 & I-E & FOZ & 08:39:28.5 & 1.1  & $3.769\cdot 10^{-3}$ \\
		2016-02-24 & I-G & OPD & 01:53:25.5 & 1.1  & $3.508\cdot 10^{-3}$ \\
		& & FEG & 01:53:27.3 & 4.0    & $1.276\cdot 10^{-2}$ \\
		2016-02-25 & I-E & GOA & 23:55:58.2 & 2.4  & $7.070\cdot 10^{-3}$  \\
		2016-03-04 & I-E & GOA & 02:09:59.3 & 2.3  & $4.095\cdot 10^{-3}$ \\
		2016-04-02 & I-E & FOZ & 05:45:57.1 & 2.2  & $3.206\cdot 10^{-3}$ \\
		& & OPD & 05:46:03.2 & 2.5  & $3.643\cdot 10^{-3}$ \\
		& & FEG & 05:45:59.1 & 3.8  & $5.538\cdot 10^{-3}$ \\
		2016-04-02 & I-C & OPD & 23:24:20.4 & 1.2  & $1.846\cdot 10^{-3}$ \\
		& & FOZ & 23:24:22.4 & 1.4  & $2.153\cdot 10^{-3}$ \\
		& & FEG & 23:24:22.3 & 3.5  & $5.383\cdot 10^{-3}$ \\
		2016-04-12 & I-C & OPD & 04:35:29.7 & 8.9  & $1.631\cdot 10^{-2}$ \\
		& & FOZ & 04:35:31.1 & 1.1  & $2.016\cdot 10^{-3}$ \\
		& & FEG & 04:35:29.1 & 2.5  & $4.581\cdot 10^{-3}$ \\
		2016-04-12 & I-E & FOZ & 04:45:49.0 & 10.1 & $1.167\cdot 10^{-3}$ \\
		2016-04-12 & E-C & FOZ & 05:01:34.6 & 1.9  & $3.337\cdot 10^{-3}$ \\
		& & FEG & 05:01:36.1 & 4.2  & $7.376\cdot 10^{-3}$ \\
		2016-04-12 & I-E & OPD & 21:17:16.2 & 0.8  & $1.653\cdot 10^{-3}$ \\
		2016-04-19 & I-E & OPD & 23:35:15.3 & 1.0    & $2.456\cdot 10^{-3}$ \\
		& & FOZ & 23:35:14.2 & 2.1  & $5.158\cdot 10^{-3}$ \\
		& & GOA & 23:35:13.3 & 2.2  & $5.404\cdot 10^{-3}$ \\
		& & UTF & 23:35:15.2 & 3.2  & $7.860\cdot 10^{-3}$  \\
		& & OHP & 23:35:13.9 & 1.5  & $3.686\cdot 10^{-3}$ \\
		2016-04-20 & E-C & OHP & 20:15:57.8 & 1.8  & $2.474\cdot 10^{-3}$ \\
		2016-04-24 & I-G & OPD & 22:35:12.0 & 0.5  & $2.609\cdot 10^{-3}$ \\
		& & UTF & 22:35:13.1 & 2.6  & $1.357\cdot 10^{-2}$ \\
		2016-04-29 & I-G & OPD & 00:32:28.1 & 2.4  & $1.023\cdot 10^{-2}$ \\
		& & UTF & 00:32:28.6 & 4.2  & $1.790\cdot 10^{-2}$  \\
		2016-05-02 & I-G & OPD & 01:08:50.3 & 1.5  & $7.411\cdot 10^{-3}$ \\
		& & FOZ & 01:08:50.7 & 2.3  & $1.136\cdot 10^{-2}$ \\
		& & FEG & 01:08:49.1 & 1.8  & $8.893\cdot 10^{-3}$ \\
		& & UTF & 01:08:51.1 & 4.5  & $2.223\cdot 10^{-2}$ \\
		2016-05-03 & E-G & OPD & 01:04:55.4 & 1.3  & $2.391\cdot 10^{-3}$ \\
		& & UTF & 01:04:55.5 & 1.9  & $3.494\cdot 10^{-3}$ \\
		2016-05-06 & E-C & OPD & 00:59:06.8 & 6.5  & $7.955\cdot 10^{-3}$ \\
		2016-05-19 & E-G & FOZ & 22:52:31.9 & 1.0    & $3.485\cdot 10^{-3}$ \\
		2016-05-27 & E-G & FEG & 02:00:21.8 & 5.5  & $1.853\cdot 10^{-2}$ \\
		2016-06-17 & I-E & OPD & 00:48:02.9 & 1.3  & $1.059\cdot 10^{-2}$ \\
		& & FEG & 00:48:07.0 & 4.8  & $3.910\cdot 10^{-2}$  \\
		2016-06-28 & I-G & OPD & 23:58:57.1 & 1.4  & $3.857\cdot 10^{-3}$ \\
		& & FEG & 23:58:59.0 & 1.1  & $3.031\cdot 10^{-3}$ \\
		2016-06-28 & I-E & OPD & 22:36:02.2 & 0.5  & $2.567\cdot 10^{-3}$ \\
		& & FEG & 22:36:02.9 & 1.2  & $6.160\cdot 10^{-3}$  \\
		2016-07-08 & E-G & OPD & 21:51:35.5 & 0.6  & $1.445\cdot 10^{-3}$ \\
		& & FEG & 21:51:32.6 & 3.3  & $7.946\cdot 10^{-3}$ \\ \hline                  
	\end{tabular}
	\tablefoot{The alternative observables' weights are provided in mas/s for the mutual approximations reported in \cite{morgado2019approx} for the year 2016. I-E, I-G, I-C, E-G and G-C refer to mutual approximations between Io and Europa, Io and Ganymede, Io and Callisto, Europa and Ganymede, Europa and Callisto, and Ganymede and Callisto, respectively. This table is adapted from \cite{morgado2019approx}.}
\end{table*}

\begin{table*}[h]
	\caption{Appropriate weights for alternative observables, for the mutual approximations observed in 2017.}
	\label{tab:weightsPastObservations2017}
	\centering
	\begin{tabular}{cccccc} \hline \hline \noalign{\vskip 0.03in}
		Date & Event & Station & $t_c$ (UTC) & $\sigma(t_c)$ [s] & $\sigma_{\mathrm{alt.}}$ [mas/s]\\ \hline \noalign{\vskip 0.03in}
		2017-02-07 & I-E & FOZ & 04:36:54.1 & 1.0    & $3.525\cdot 10^{-3}$ \\
		2017-02-26 & I-E & FOZ & 04:32:43.5 & 1.3  & $4.368\cdot 10^{-3}$ \\
		2017-02-27 & I-G & FOZ & 03:36:51.3 & 1.1  & $1.609\cdot 10^{-3}$ \\
		2017-03-07 & I-G & FOZ & 03:00:44.4 & 32.9 & $2.126\cdot 10^{-3}$ \\
		2017-03-14 & I-G & FOZ & 07:19:33.8 & 1.1  & $6.109\cdot 10^{-4}$ \\
		2017-04-04 & I-E & OHP & 20:43:34.4 & 0.7  & $3.227\cdot 10^{-3}$ \\
		2017-04-06 & I-E & FEG & 03:46:43.1 & 2.2  & $6.249\cdot 10^{-3}$ \\
		2017-04-08 & E-G & FOZ & 01:52:40.5 & 1.0    & $1.924\cdot 10^{-3}$ \\
		2017-04-13 & I-E & FOZ & 05:49:28.3 & 1.0    & $2.712\cdot 10^{-3}$ \\
		2017-05-06 & I-G & GOA & 02:16:30.2 & 1.7  & $4.460\cdot 10^{-3}$  \\
		2017-05-08 & I-E & FOZ & 01:11:26.5 & 1.0    & $2.124\cdot 10^{-3}$ \\
		2017-05-13 & I-G & FOZ & 04:47:32.1 & 1.0    & $2.418\cdot 10^{-3}$ \\
		2017-05-15 & I-E & FEG & 03:23:43.1 & 1.7  & $3.237\cdot 10^{-3}$ \\
		2017-05-31 & E-G & FEG & 22:30:36.2 & 27.9 & $9.891\cdot 10^{-4}$ \\
		2017-06-08 & I-E & FEG & 23:48:57.1 & 7.5  & $6.752\cdot 10^{-3}$ \\
		& & GOA & 23:48:58.1 & 1.8  & $1.621\cdot 10^{-3}$ \\
		2017-06-23 & I-E & FOZ & 23:17:09.0 & 1.1  & $8.644\cdot 10^{-4}$ \\
		& & GOA & 23:17:07.7 & 1.9  & $1.493\cdot 10^{-3}$ \\
		2017-07-06 & E-G & FOZ & 22:58:42.6 & 1.4  & $8.937\cdot 10^{-4}$ \\
		& & FEG & 22:58:41.1 & 19.4 & $1.238\cdot 10^{-2}$ \\
		2017-07-25 & I-E & FOZ & 22:40:24.8 & 1.2  & $2.166\cdot 10^{-3}$ \\
		& & FEG & 22:40:21.3 & 3.3  & $5.957\cdot 10^{-3}$ \\
		2017-08-02 & G-C & FEG & 23:38:20.0 & 7.7  & $4.172\cdot 10^{-3}$ \\
		2017-08-10 & E-C & FOZ & 23:41:23.6 & 48.2 & $1.680\cdot 10^{-3}$  \\
		2017-08-24 & I-G & FEG & 22:35:37.6 & 6.6  & $3.915\cdot 10^{-3}$ \\\hline
	\end{tabular}
	\tablefoot{The alternative observables' weights are provided in mas/s for the mutual approximations reported in \cite{morgado2019approx} for the year 2017. I-E, I-G, I-C, E-G and G-C refer to mutual approximations between Io and Europa, Io and Ganymede, Io and Callisto, Europa and Ganymede, Europa and Callisto, and Ganymede and Callisto, respectively. This table is adapted from \cite{morgado2019approx}.}
\end{table*}

\begin{table*}[h]
	\caption{Appropriate weights for alternative observables, for the mutual approximations observed in 2018.}
	\label{tab:weightsPastObservations2018}
	\centering
	\begin{tabular}{cccccc} \hline \hline \noalign{\vskip 0.03in}
		Date & Event & Station & $t_c$ (UTC) & $\sigma(t_c)$ [s] & $\sigma_{\mathrm{alt.}}$ [mas/s]\\ \hline \noalign{\vskip 0.03in}
		2018-03-05 & I-E & FOZ & 05:10:29.7 & 0.6  & $1.378\cdot 10^{-3}$ \\
		2018-03-11 & I-G & OPD & 05:40:46.7 & 1.8  & $6.834\cdot 10^{-4}$ \\
		& & FOZ & 05:40:47.0 & 2.0    & $7.594\cdot 10^{-4}$ \\
		2018-03-12 & I-E & OPD & 07:20:57.6 & 0.5  & $1.236\cdot 10^{-3}$ \\
		& & FOZ & 07:20:58.8 & 1.4  & $3.461\cdot 10^{-3}$ \\
		2018-03-17 & I-E & FOZ & 03:15:03.2 & 0.8  & $2.898\cdot 10^{-3}$ \\
		2018-03-24 & I-E & FOZ & 05:18:47.9 & 0.7  & $2.527\cdot 10^{-3}$ \\
		2018-04-06 & I-E & OPD & 02:40:32.0 & 1.2  & $3.641\cdot 10^{-3}$ \\
		& & FOZ & 02:40:31.4 & 1.0    & $3.034\cdot 10^{-3}$ \\
		2018-06-11 & E-G & FEG & 23:03:46.0 & 1.8  & $2.885\cdot 10^{-3}$ \\
		& & GOA & 23:03:45.1 & 1.2  & $1.923\cdot 10^{-3}$ \\
		2018-06-19 & E-G & FOZ & 01:55:19.9 & 1.1  & $1.785\cdot 10^{-3}$ \\
		2018-06-22 & I-G & FEG & 02:17:09.5 & 7.2  & $1.005\cdot 10^{-3}$ \\
		& & OPD & 02:17:12.6 & 4.5  & $6.282\cdot 10^{-4}$ \\
		& & FOZ & 02:17:12.5 & 5.6  & $7.818\cdot 10^{-4}$ \\
		& & GOA & 02:17:09.9 & 6.5  & $9.074\cdot 10^{-4}$ \\
		2018-06-23 & I-E & FOZ & 00:40:47.4 & 1.1  & $4.486\cdot 10^{-3}$ \\
		2018-07-07 & I-G & OPD & 00:30:56.8 & 1.1  & $1.829\cdot 10^{-3}$ \\
		& & FEG & 00:30:57.0 & 2.2  & $3.658\cdot 10^{-3}$ \\
		2018-07-11 & E-C & OPD & 22:48:02.8 & 1.4  & $1.010\cdot 10^{-3}$  \\
		2018-07-12 & I-E & FEG & 01:07:36.3 & 2.5  & $5.285\cdot 10^{-3}$ \\
		& & OPD & 01:07:37.4 & 1.0    & $2.114\cdot 10^{-3}$ \\
		2018-07-13 & E-G & OPD & 02:01:30.9 & 1.1  & $9.424\cdot 10^{-4}$ \\
		& & FEG & 02:01:29.9 & 5.4  & $4.626\cdot 10^{-3}$ \\
		2018-07-19 & I-C & OPD & 01:52:08.6 & 1.9  & $1.502\cdot 10^{-3}$ \\
		& & FOZ & 01:52:09.3 & 2.1  & $1.661\cdot 10^{-3}$ \\
		2018-08-07 & E-G & OPD & 23:15:18.8 & 1.3  & $2.118\cdot 10^{-3}$ \\
		2018-08-12 & I-E & OPD & 23:54:58.4 & 1.1  & $1.186\cdot 10^{-3}$ \\
		& & FOZ & 23:54:58.5 & 1.2  & $1.294\cdot 10^{-3}$ \\ \hline                    
	\end{tabular}
	\tablefoot{The alternative observables' weights are provided in mas/s for the mutual approximations reported in \cite{morgado2019approx} for the year 2018. I-E, I-G, I-C, E-G and G-C refer to mutual approximations between Io and Europa, Io and Ganymede, Io and Callisto, Europa and Ganymede, Europa and Callisto, and Ganymede and Callisto, respectively. This table is adapted from \cite{morgado2019approx}.}
\end{table*}

\end{document}